\documentclass[twocolumn,showpacs,amsmath,amssymb,aps,prb,superscriptaddress,a4paper]{revtex4-1}

\usepackage{chemformula} % Formula subscripts using \ch{}
\usepackage[T1]{fontenc} % Use modern font encodings
\usepackage[utf8]{inputenc}
\usepackage[frenchb, english]{babel}
\usepackage{amsmath}
\usepackage{amssymb}
\usepackage{amsfonts}
\usepackage{braket}
\usepackage{lmodern}
\usepackage{mathptmx}
\begin{document}

\title{A comprehensive model of the optical spectra of carbon nanotubes on substrate by polarized microscopy}

\author{L\'{e}onard Monniello}

\author{Huy-Nam Tran}

\author{Rémy Vialla}

\author{Guillaume Prévot}

\author{Said Tahir}

\author{Thierry Michel}

\author{Vincent Jourdain}
\email{vincent.jourdain@umontpellier.fr}

\affiliation{Laboratoire Charles Coulomb (L2C), Université de Montpellier, CNRS, Montpellier, France}


\begin{abstract}
  Polarized optical microscopy and spectroscopy are progressively becoming key methods for the high-throughput characterization of individual carbon nanotubes (CNTs) and other one-dimensional nanostructures, on substrate and in devices. The optical response of CNTs on substrate in cross polarization experiments is usually limited by the polarization conservation of the optical elements in the experimental setup. We developed a theoretical model taking into account the depolarization by the setup and the optical response of the substrate. We show that proper modelization of the experimental data requires to take into account both non-coherent and coherent light depolarization by the optical elements. We also show how the nanotube signal can be decoupled from the complex reflection factor of the anti-reflection substrate which is commonly used to enhance the optical contrast. Finally, we describe an experimental protocol to extract the depolarization parameters and the complex nanotube susceptibility, and how it can improve the chirality assignment of individual carbon nanotubes in complex cases.
\end{abstract}

\maketitle

\section{Introduction}

One of the most fascinating aspects of single-walled carbon nanotubes (SWCNTs) is the extreme dependence of their properties on their crystalline structure and their environment. However, this remarkable feature is a two-sided coin: it opens up a wealth of new phenomena and applications \cite{Zhang2017,Volder2013} but is problematic when specific and well-controlled properties are required as for electronic applications \cite{Jariwala2013}. This problem is even more acute when dealing with individual SWCNTs because measuring their structure- or environment-dependent signatures is usually long and difficult, especially when studied \textit{in situ}\cite{Rao2012, Picher2009} or \textit{in operando} in devices. As a direct method of structural characterization, electron diffraction should be the method of choice for determining the structure of individual nanotubes \cite{Ghedjatti2017} but is severely limited by the requirement of using suspended SWCNTs, and by the cost and complexity of transmission electron microscopes. In practice, Raman spectroscopy is the most popular method because it is simple to perform, is appropriate for both metallic and semiconducting SWCNTs, provides information-rich spectra and is widely available. However, a main drawback is the trial-and-error approach required to determine a laser energy in resonance with an optical transition of the studied nanotube. By measuring the optical transition energies of individual SWCNTs, resonant Rayleigh scattering represents a high-throughput method of structural identification \cite{Sfeir2004} but is usually restricted to suspended nanotubes because of the intense reflection or scattering from the substrate. Recently, novel optical methods have been developed to directly measure the optical features of individual nanotubes directly on substrates.\cite{Joh2010,Wu2015,Havener2011} Among them, polarized optical spectroscopy \cite{Lefebvre2011,Liu2013,Deng2016} is of particular interest since it imposes no additional constraint (\textit{e.g.} deposited layer, liquid environment) on the sample.

\begin{figure}
\begin{center}
\includegraphics[scale=0.5]{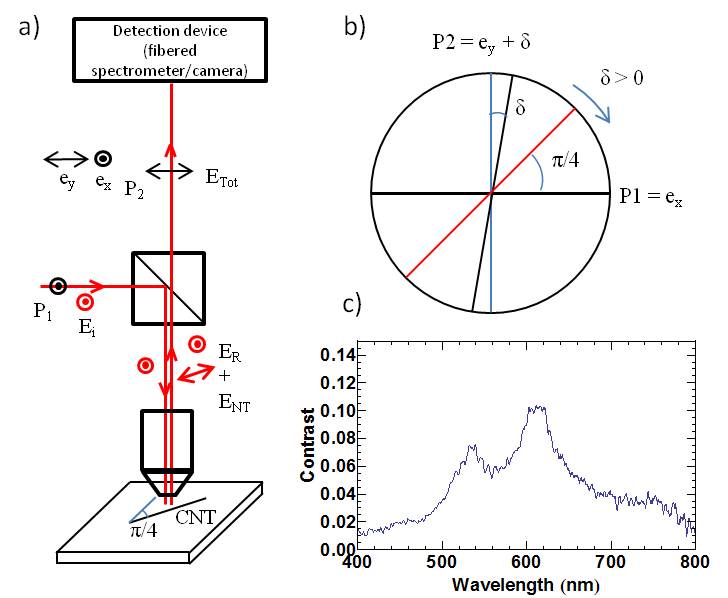}
\end{center}
\caption{a) Schema of the experiment. b) Cross polarizers configuration. c) Typical polarized optical spectrum of a CNT on substrate}
\label{figure1}
\end{figure}

The method relies on the strong polarization effect of SWCNTs along their main axis. By setting the SWCNT between two crossed polarizers, at an angle of 45\degres \space with the main axes of the polarizers, the intense substrate signal which keeps the incident polarization can be reduced by several orders of magnitude, thus making the weak nanotube signal observable (see Figure \ref{figure1}). The theoretical treatment is however complicated by several factors. In their seminal article on suspended SWCNTs, Lefebvre and Finnie first took into account the interference between the nanotube signal and the transmitted light, and the imperfection of the polarization system. \cite{Lefebvre2011} Extending the method to SWCNTs on substrate in a reflection configuration, Liu \textit{et al.} considered the contribution of the reflection factor of the substrate in their model with the assumption of a perfect polarization system.\cite{Liu2013} More recently, Deng \textit{et al.} extended the model by considering an imperfect polarization system to account for the finite contrast at small angles of crossed polarization.\cite{Deng2016} However, even this latest model has limitations preventing it to reproduce all the experimental features and extract all the information available on the nanotube. First, it only considers uncoherent depolarization by the optical system and neglects the possibility of coherent depolarization which is needed to account for interferential features. Second, the complex nature of the reflection factor is not developed, which is especially important in the case of Si substrates with a thermal oxide layer (SiO$_2$) acting as an anti-reflection coating. Such considerations are notably important because they would theoretically allow one to measure both the real and imaginary parts of the nanotube susceptibility, and not only the imaginary one as commonly done today. Very recently, F. Yao \textit{et al} reported a method to measure the complex susceptibility on individual nanotubes using elliptical polarization {\cite{Yao2018} but this method cannot be used on the SiO$_2$/Si substrates which are commonly used for nanotube devices, due to the complex nature of the reflection coefficients.
Here, we describe a model including all the above-mentioned contributions and show that it allows one to nicely reproduce the experimental spectra of SWCNTs on different substrates. We also report methods to correct the spectra from the imperfections of the optical system and to extract the real and imaginary parts of the nanotube susceptibility.

\section{Experimental setup}

The experimental setup is shown in figure \ref{figure1}. The light source (Fianium SC-400-4, 2 ps pulses, 40 MHz) is polarized by a Glan-laser polarizer ($P_1$) along the axis $e_x$. A 80/20 beamsplitter allows illuminating the sample and collecting the signal. Supported  nanotubes, which are oriented at $45\degres$ from $e_x$ (figure \ref{figure1}b), are illuminated by a microscope objective (NA 0.8, WD 1.4 mm) chosen for its polarization conservation properties. A second Glan polarizer ($P_2$) analyses the light from the sample which is then detected by either a camera or a fibered spectrometer ($P_1$ and $P_2$ have shown extinction ratio above $10^7$ in optimal conditions). The scattered field from the nanotube ($E_{NT}$) is polarized along the nanotube axis while the field reflected by the substrate ($E_r$) remains polarized along the $e_x$ direction. The diameter of the incident beam is controlled by a lens based beam reducer and an iris to improve the polarization conservation of the microscope objective by reducing the numerical aperture (NA). The typical diameter used is 2 mm for an effective NA of the incident beam of 0.4.
With $E_{NT}/E_R$ typically being in the order of $10^{-4}$, nanotubes on substrate can be observed either by cutting the reflected intensity by a factor $10^8$ or through interference between the reflected field and the scattered field. To detect the latter with a contrast of around $1\%$, one needs an extinction ratio in the order of $10^{-4}$ in intensity, \textit{i.e} $10^{-2}$ in field. The interference signal can then be observed through a homodyne detection of the intensity on the nanotube ($I$) and out of the nanotube ($I_0$) which gives the contrast $C=(I-I_0)/I_0$. To account for temporal fluctuations of the laser intensity, the sample position is modulated between the two positions and intensities are summed. Figure \ref{figure1}c shows a typical contrast spectrum of a SWNT displaying two broad features at 540 nm and 620 nm.

\section{Modelisation}

We now discuss the possible models to describe the experimental spectra. First, we focus on the ideal case where the polarization of the light is conserved throughout the entire setup, as already described in reference \cite{Liu2013}. Second, we discuss the limitations of this model and how to make it more realistic by introducing depolarization by the optical elements. Finally, we model the influence of the substrate, notably in the case of an anti-reflection layer with complex reflection and amplification factors. All angles and axes orientations used in the proposed models are defined in figure \ref{figure1}b.

\subsection{Model for ideal optics.}

First, let us consider the ideal case where the optical elements do not depolarize the light. The incident field $\overrightarrow{E_i}$ can be written as:

\begin{equation}
\overrightarrow{E_i}=|E_0|e^{i\phi}\overrightarrow{e_x}
\end{equation}

where $|E_0|$ is the amplitude of the incident field and $\phi$ is the Gouy phase\cite{Gouy}, the additional phase occuring in the propagation of a focused Gaussian beam. $\phi$ is equal to $\pi/2$ between the far field and the center of the waist, and takes another $\pi/2$ when reaching the far field. On the sample, this field is either reflected by the substrate or scattered after interacting with the CNT. The reflection keeps the polarization of the incident beam and follows this expression:

\begin{equation}
\overrightarrow{E_R}=r\overrightarrow{E_i}e^{i\phi}=r|E_0|e^{i\pi}\overrightarrow{e_x}
\end{equation}

where $r$ is the amplitude reflection coefficient ($r^2$ is the intensity reflection coefficient). Here, the Gouy phase for the reflected beam is taken equal to $\pi$ since the detection is done in the far field. The field scattered by the CNT is considered totally repolarized along the CNT axis, and writes as:

\begin{equation}
\overrightarrow{E_{NT}} = A\frac{(1+r)^2|E_0|e^{i\phi}}{2}\left(\overrightarrow{e_x}+\overrightarrow{e_y}\right)
\end{equation}

where $A$ is a complex constant which is proportional to the nanotube susceptibility $\chi$ and depends on geometrical parameters (\textit{e.g.} the position of the CNT in the illuminated area or the collection efficiency). The $(1+r)$ factor denotes the fact that the nanotube sees both the incident light and the reflected light. This factor is squared because this effect acts for both absorption and scattering. $\phi$ is the Gouy phase acquired by the field before exciting the CNT which depends on the position of the CNT within the waist: if the CNT is exactly at the beam waist, $\phi=\pi/2$. The scattered light from the CNT does not gain additional phase when going in the far field. Both fields are collected by the microscope objective and analysed through the second polarizer. The intensity $I$ measured on the camera or the spectrometer is given by the expression:

\begin{equation}
I=|E_{tot}|^2= |\left(\overrightarrow{E_R}+\overrightarrow{E_{NT}}\right).\left(\sin(\delta)\overrightarrow{e_x}+\cos(\delta)\overrightarrow{e_y}\right)|^2
\end{equation}

where both fields are summed and projected on the main axis of the second polarizer which makes a small angle $\delta$ with the $e_y$ direction. This intensity is compared to the reference intensity $I_0$ reflected by the substrate, away from the CNT:

\begin{equation}
I_0=|\overrightarrow{E_R}.\left(\sin(\delta)\overrightarrow{e_x}+\cos(\delta)\overrightarrow{e_y}\right)|^2
\end{equation}

The contrast between the two signals is given by (see SI section 1A for the calculation details):

\begin{equation}
\label{contrast_ideal}
C = \frac{I - I_0}{I_0} \approx \frac{(1+r)^4|A|^2}{4r^2\delta^2}+\frac{(1+r)^2Re(e^{i(\pi-\phi)}A^*)}{r\delta}
\end{equation}

The first term of this equation is related to scattering and shows what would be measured in a Rayleigh scattering experiment. The second term describes the interference between the reflected field and the scattered field of the CNT. This term is called interference term in the following of this paper. If $\phi$ is equal to $\pi/2$, i.e the CNT is at the beam waist, then $Re(e^{i\pi-\phi}A^*)=Im(A)$ and we find the same expression as in reference \cite{Liu2013}. One can see that, if the first term is negligible, the measured contrast is directly proportional to the nanotube absorption coefficient. In this case, the sign of the contrast changes with the sign of the analyzer angle (figure \ref{figure2}, black and red spectra). However, this is only true when $\delta$ is relatively large, which means that one needs to work at an intermediate contrast value to directly extract the absorption. At smaller angles $\delta$ it is theoretically possible to extract both $Im(A)$ and $|A|^2$, and therefore the imaginary and real parts of A, but this requires at least two measurements at different $\delta$ values. \\

There are however a few issues with this ideal case model when dealing with small $\delta$ angles in the range of $\pm 5.10^{-3}$ rad. First, expression (\ref{contrast_ideal}) diverges for $\delta \rightarrow 0$ which is unphysical. Experimentally, the contrast is actually observed to decrease below a certain threshold angle value. Second, periodic oscillations experimentally appear in the contrast spectrum, which cannot be explained with this simple model. These oscillations are illustrated in figures \ref{figure2}a and b (blue spectra) which illustrate two experimental cases named cases A and B in the following.

\begin{figure}
\begin{center}
\includegraphics[scale=0.4]{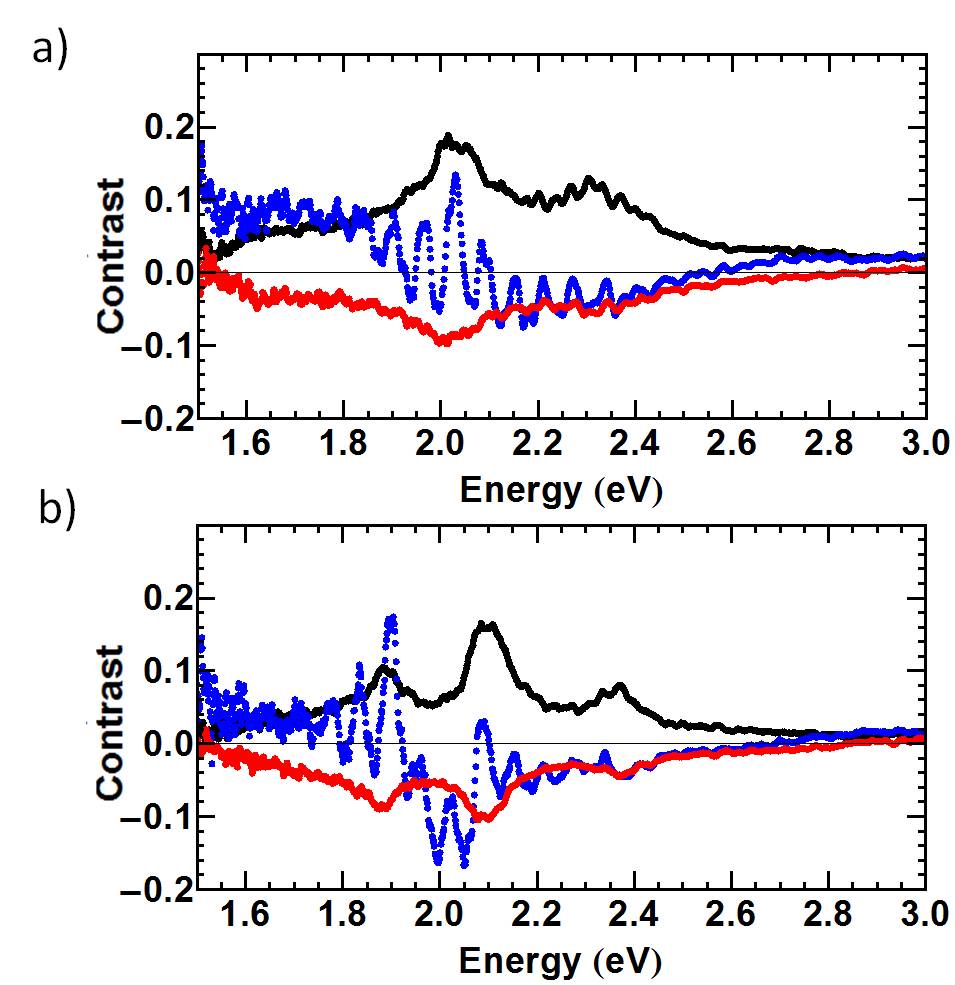}
\end{center}
\caption{Spectra of carbon nanotubes for three different angles: 10$^{-2}$ (black), -5.10$^{-3}$ (blue) and -5.10$^{-2}$ (red) rad.
a) Case A. b) Case B. In both cases, the substrate is 100 nm SiO$_2$/Si.}
\label{figure2}
\end{figure}

\subsection{Models with depolarization by the optics.}

To explain these oscillations we developed a model that considers that the polarization of the incident beam on the CNT is not perfectly linear. This effect has already been partially considered in reference \cite{Deng2016}. Here, we develop a more general model considering the two possible types of depolarization.

\begin{itemize}
\item Non-coherent depolarization. As in reference \cite{Deng2016}, we consider that the information contained in the polarization is lost. One can model the polarization at the output of the microscope objective as the sum of a linear polarization (the same as that of the input light) and a randomly polarized light which can be modeled as the sum of two orthogonal linearly-polarized lights with random phases (figure \ref{figure3}b). The light now has an elliptic polarization with one axis being the same as that of the input polarization. Because the randomly polarized light will not interfere with the incident or reflected fields, we call this case non-coherent depolarization. The portion of the input light that is depolarized along the $e_x$ axis (resp. the $e_y$ axis) is called $f_{x}$ (resp $f_y$)  and is considered to be much smaller than 1. We can now write:

\begin{equation}
\overrightarrow{E_i} = |E_i| \left(a\overrightarrow{e_x}+f_xe^{i\psi_x(t)}\overrightarrow{e_x}+f_ye^{i\psi_y(t)}\overrightarrow{e_y}\right)
\end{equation}

with $\psi_x(t)$ and  $\psi_y(t)$ the random phases of the depolarized light (see SI, part 1B  and reference \cite{Bylander} for the randomization procedure). The coefficients $a$, $f_x$ and $f_y$ follow the normalization equation $|a+f_xe^{i\psi_x(t)}|^2+|f_ye^{i\psi_y(t)}|^2=1$. As detailed in the SI, the expression of the contrast is now given by the expression:

\begin{equation}
C \approx \frac{(1+r)^4|A|^2}{4r^2\left(\delta^2+f_y^2\right)} + \frac{\delta (1+r)^2 Re(e^{i(\pi-\phi)}A^*)}{r\left(\delta^2 + f_y^2\right)}
\end{equation}

One can note that by adding depolarization, expression (8) does not diverge anymore for $\delta \rightarrow 0$. In the particular case where $\phi=\pi/2$, this expression can be simplified to write:

\begin{equation}
C \approx \frac{(1+r)^4|A|^2}{4r^2\left(\delta^2+f_y^2\right)} + \frac{\delta(1+r)^2 Im(A)}{r\left(\delta^2 + f_y^2\right)}
\end{equation}

This expression is similar to the one given in reference \cite{Deng2016} and shows that non-coherent depolarization does not change the nature of the contrast. In other words, the shape of the spectra will not change. It is therefore needed to take into account another type of depolarization to account for the periodic oscillations observed at small values of $\delta$.

\item Elliptization. Here, we consider that the optical elements act as a waveplate which changes the linear polarization into an elliptic one with a fixed phase between the two orthogonal components (figure \ref{figure3}c). The portion of the input field that is depolarized is called $f_{co}$ and is assumed to be much smaller than 1. The incident field now writes as

\begin{equation}
\overrightarrow{E_i} = |E_i|e^{i\phi}(\overrightarrow{e_x}+f_{co}e^{i\psi}\overrightarrow{e_y})
\end{equation}

where $\psi$ is the phase of the depolarized light. Here, contrary to the previous case, the coherently depolarized light can interfere with the input polarization when interacting with a CNT oriented at $45 \degres$.  As detailed in SI (part 1B), the contrast is then given by:

\begin{equation}
C= \frac{(1+r)^4|A|^2}{4r^2|\delta +f_{co} e^{i\psi}|^2}+ \frac{(1+r)^2Re\left(e^{i(\pi-\phi)}A^*(\delta +f_{co} e^{i\psi})\right)}{r|\delta +f_{co} e^{i\psi}|^2}
\label{eq_cont}
\end{equation}

By considering the particular case where $\phi=\pi/2$, one can note that A is now coupled to a complex depolarization term as can be seen in the numerator of the interference term. The complete expression can be explicited as:

\begin{equation}
\begin{aligned}
C &= \frac{(1+r)^4|A|^2}{4r^2|\delta +f_{co} e^{i\psi}|^2}\\
&+ (1+r)^2\frac{(\delta+f_{co}\cos\psi)Im(A)- f_{co}\sin\psi Re(A)}{r|\delta +f_{co} e^{i\psi}|^2}
\end{aligned}
\end{equation}

The numerator of the interference term is now composed of two contributions. The first contribution involves the nanotube absorption and is no more canceled for $\delta=0$ but for $\delta=-f_{co}\cos\psi$ which will be named $\delta_0$ in the following. The second contribution is related to $Re(A)$, which is itself related to $Im(A)$ through Kramers-Kronig relations\cite{Yao2018}:

\begin{equation}
\label{KK}
Re(\chi(\omega))=\frac{2}{\pi}\int_0^{\infty}\frac{\omega'Im(\chi\omega'))}{\omega'^2-\omega^2}d\omega'
\end{equation}
\noindent where $\omega$ is the energy. The second term in the expression of C can change drastically the shape of the contrast spectrum, providing that $\sin\psi$ is comparable to $\cos\psi$. Let us define $f=f_{co}sin\psi$ and the expression can be written as :

\begin{equation}
\begin{aligned}
C &= \frac{(1+r)^4|A|^2}{4r^2|\delta - \delta_0 + i f|^2}\\
&+ (1+r)^2\frac{(\delta-\delta_0)Im(A)- f Re(A)}{r|\delta -\delta_0 + i f|^2}
\end{aligned}
\end{equation}

This expression shows that in the specific case where $f=0$, \textit{i.e.} $\psi=0$ (or $\pi$), the contrast diverges for $\delta=\delta_0$: the two components of the polarization are in phase (or in opposition of phase) and the resulting polarization is linear. In other words, to fully model the physics behind the contrast spectrum, one must take into account both types of depolarization, coherent and non coherent, and the most general expression of the contrast is then given by :

\begin{equation}
\label{eqcont3}
\begin{aligned}
C &= \frac{(1+r)^4|A|^2}{4r^2(|\delta -\delta_0 + i f|^2+f_{y}^2)}\\
&+ (1+r)^2\frac{Re\left(e^{\pi-i\phi}A^*(\delta-\delta_0+i f)\right)}{r(|\delta -\delta_0 + i f|^2+f_{y}^2)}
\end{aligned}
\end{equation}

This expression shows the importance of knowing the type of depolarization involved in the experiment as the numerator and denominator do not have common factors. It also shows that both non-coherent and coherent depolarizations will have a coupled effect (in the denominator) and cannot be easily separated if both occur in the experiment.
\end{itemize}

\begin{figure}
\begin{center}
\includegraphics[scale=0.5]{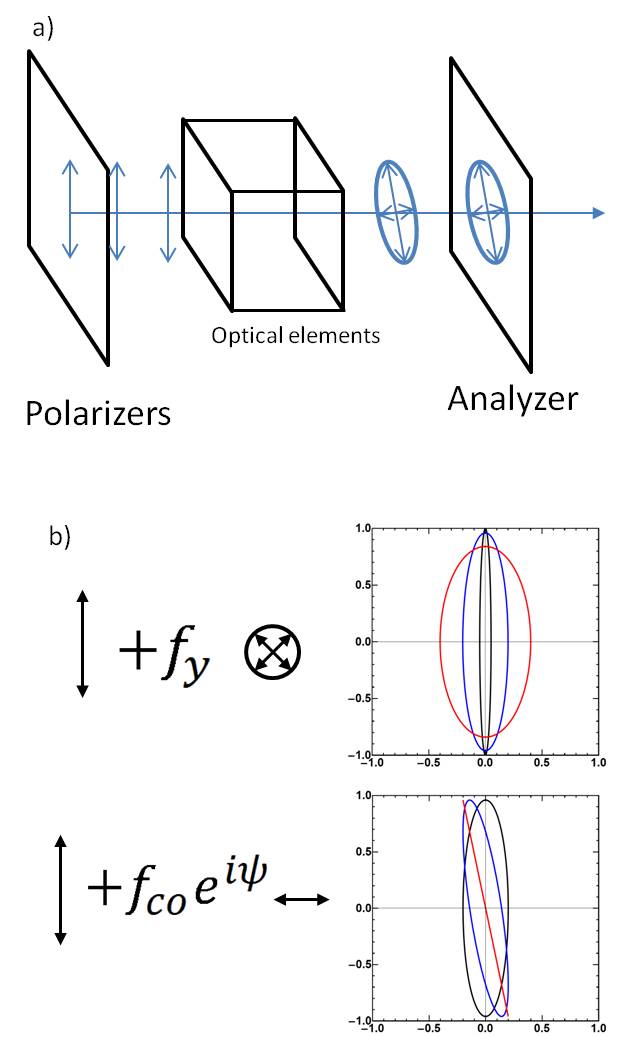}
\end{center}
\caption{a) Illustration of the depolarization by the optical elements between the polarizer and the analyzer. The polarization can be decomposed along two orthogonal axes.
	 b) Top: non-coherent depolarization. The information of the phase is lost on the perpendicular axis. Black (resp. blue, red) is the polarization with $f_{y}$ =0.05 (resp. 0.2, 0.4).
	      Bottom : coherent depolarization. The phase of the perpendicular polarization is shifted. Black (resp. blue, red) is the polarization with $f_{co}$ = 0.2 and $\psi$=0 (resp. $\pi$/4, $\pi$/2)
} 
\label{figure3}
\end{figure}

\subsection{Substrate model.}

To fully model the experimental spectra, one must also take into account the effect of the substrate. In the previous part, we considered a mirror substrate with a real reflection coefficient, e.g. a silica substrate or a silicon substrate without oxide layer. In such a case, $r$ is negative and the smaller $|r|$ is, the higher the contrast will be by both influencing the denominator term ($r$) and the numerator term ($1+r$) (see for instance eq 6). In this part we will now consider the case of multilayer substrates such as SiO$_2$/Si which leads to multiple interfering reflections. These interferences lead to two distinct reflection coefficients. The first, $R_{E}$ is the field effective reflection coefficient which is equal to $r$ on a mirror substrate. The second, $R_A$, is the effective coefficient characterizing the portion of the reflected field that interacts with the CNT (i.e $ (1+r)$ in the mirror substrate case). The latter can be seen as an amplification factor of the field interacting with the nanotube and is called "amplification factor" in the following of the paper. The interferences in the oxide layer is a well known phenomenon (developed in SI part 2), which causes the two reflection coefficients to become complex functions of the wavelength:

\begin{equation}
R_{E}=-\frac{R_{01}+R_{12}e^{i\theta}}{1+R_{12}R_{01}e^{i\theta}}
\end{equation}

and 

\begin{equation}
R_{A}= 1-\frac{(1-R_{01}^2)R_{12}e^{i\theta}}{1+R_{01}R_{12}e^{i\theta}}
\end{equation}

where $R_{01}$ (resp. $R_{12}$) is the reflection coefficient at the air/SiO$_2$ interface (resp. SiO$_2$/Si interface) and $\theta=4i\pi d/\lambda$ is the dephasing due to the travel of the beam inside the oxide layer of thickness $d$. The real and imaginary parts of the effective reflection coefficient are plotted as a function of the wavelength in figure \ref{figure4}a, for a 100 nm SiO$_2$ thickness. The reflection coefficients of silica and Si are also shown for comparison. The intensity reflection coefficient, $|R_E|^2$ is plotted in figure \ref{figure4}b for different SiO$_2$ thicknesses (100, 300 and 500 nm). The amplification factor is shown in figure \ref{figure4}c and the intensity amplification factor $|R_A|^2$ is shown in figure \ref{figure4}d. One can observe that the behavior of the amplification factor with respect to the energy matches that of the reflection factor with opposite phase. The calculation of the contrast in the ideal case without depolarization gives a new expression (see SI part 1B for details):

\begin{equation}
C = \frac{|A|^2|R_A|^4}{4|R_{E}|^2\delta^2}+\frac{Re\left(e^{i(\pi-\phi)} A^*(R_A)^2R_{E}^*\right)}{|R_{E}|^2\delta}
\label{eq_cont2}
\end{equation}

where one can see that the amplification factor only appears in the numerator. Its effect will add up to the one of the reflection coefficient, increasing further the contrast. Figure \ref{figure4}e and f illustrate the effect of the oxide thickness on the spectra. For illustration purposes, we added four CNT excitonic transitions\cite{Berciaud2010,Tran2016} of equal strength at different energies together with a non resonant susceptibility term (i.e a complex constant $A_0$):

\begin{equation}
A(\lambda) = \frac{C_0}{(\omega_0-\omega)-i\frac{\Gamma}{2}}+A_0
\end{equation}

where $\omega_0$ is the central frequency of a given excitonic transition, $\Gamma$ its width and $C_0$ its amplitude. One can observe that the oxide thickness influences the profile of the background as well as the contrast of the optical transitions. The SiO$_2$ coating on Si multiplies the contrast of nanotubes by up to a factor 50 compared with silicon and by up to a factor 3 compared with fused silica (figure 4f, top). This model shows also that the thickness of the coating can be adjusted to match the energy range of the optical transitions studied (figure 4e), 80 to 100 nm giving the largest range across the visible. Figure \ref{figure4}f also shows that reaching higher orders of interference via a thicker oxide layer does not provides higher contrast but makes the energy range of amplification more narrow. Experimental validation of this model has been performed on 100 nm, 300 nm and 500 nm oxide thickness substrates as shown in SI (part 2).\\

Finally, taking into account the effect of the substrate and both types of depolarization, the contrast writes as

\begin{equation}
\label{eqfinale}
\begin{aligned}
C &= \frac{|R_A|^4|A|^2}{4|R_{E}|^2(|\delta -\delta_0 + i f|^2+f_{y}^2)}\\
&+ \frac{Re\left(e^{i\phi}A^*R_A^2R_{E}^*(\delta-\delta_0+i f)\right)}{|R_{E}|^2(|\delta -\delta_0 + i f|^2+f_{y}^2)}
\end{aligned}
\end{equation}

From this expression, one can choose to use the various models previously presented. For instance, setting $f=0$ is equivalent to have a fully non-coherent depolarization. Though this expression presents many parameters, we will see in the next part that some of them can be fixed or obtained through independent measurements.\\
Importantly this final expression shows that extracting $Im(A)$ and $Re(A)$ is complexified by the coupling between $A$, $R_E$, $R_A$ and the depolarization ($\delta_0$ and $f$). This prevents a simple mathematical treatment such as dividing the contrast by the optical response of the substrate. Determining the depolarization parameters is therefore a key step to extract $Im(A)$ and $Re(A)$.

\begin{figure}
\begin{center}
\includegraphics[scale=0.4]{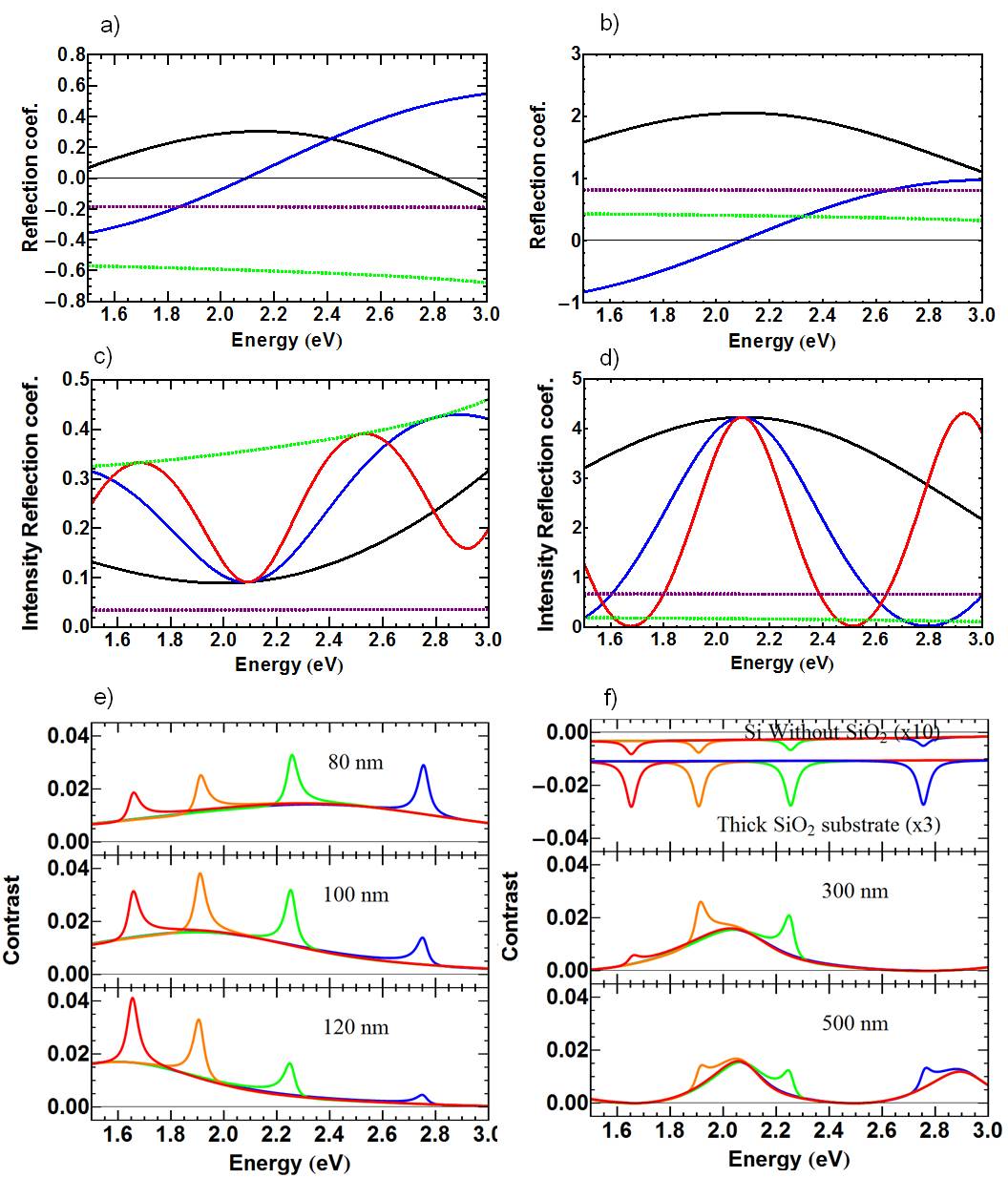}
\end{center}
\caption{a) (resp. b) Evolution of the real (black line) and imaginary (blue line) part of the field reflection coefficient $R_E$ (resp. amplification factor $R_A$). Green and purple dashed lines represent the coefficients of Si and SiO$_2$ substrates as references. c) (resp. d) Intensity reflection coefficient $|R_E|^2$ (resp. amplification factor $|R_A|^2$) for three SiO$_2$ thicknesses: 100 nm (black), 300 nm (blue) and 500 nm (red).  Green and purple dashed lines represent the coefficients of Si and SiO$_2$ substrates as references. e-f) Contrast obtained on different substrates considering four excitonic transitions (eq. 19). 
}
\label{figure4}
\end{figure}

\section{Fitting, interpretation and exploitation of experimental spectra}

\begin{figure*}
\begin{center}
\includegraphics[scale=0.6]{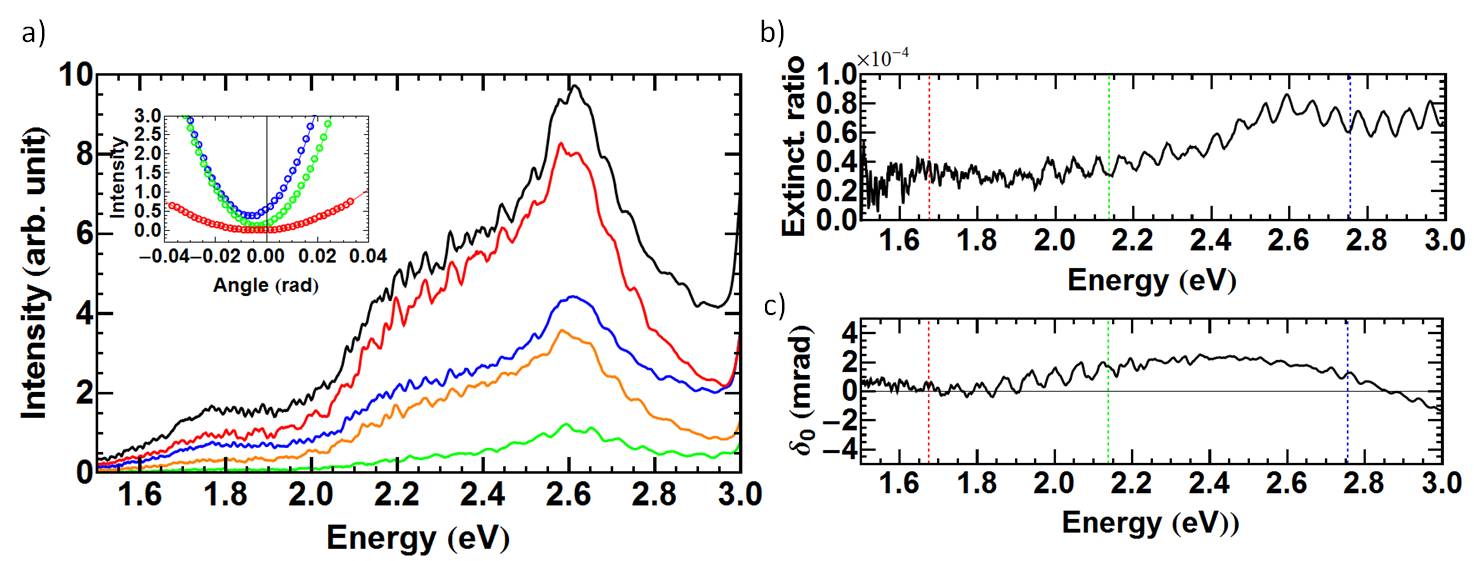}
\end{center}
\caption{a) Raw spectrum on 100 nm SiO$_2$/Si substrate for different angles $\delta$: 0.05 rad (black), 0.02 rad (blue), 0.005 rad (green), -0.02 rad (orange) -0.05 rad (red).  Inset : intensity as a function of the angle for 3 wavelengths (450 nm, 580 nm and 740 nm), fitted by eq. 20  b) and c) Parameters extracted from the fitted curves: b) extinction ratio $|f_t|^2$. c) tilt angle $\delta_0$. Vertical dotted lines show the energies used in inset.} 
\label{figure5}
\end{figure*}

As shown in the last two parts, the expression of the contrast is strongly modified by the depolarization by the optics and by a multi-layered substrate. To extract all possible information from the spectra, it is therefore required to know these parameters. Although the substrate is usually known well enough to be considered as a fixed parameter, it is more difficult to access the depolarization parameters. The latter ones will depend on the optical elements of the experimental setup : polarizers, beam splitter and objective.  It is well known that the objective is usually the most critical optical element in a polarization microscope and the injection of the beam in the objective in terms of position, angle and width is a key aspect for polarization conservation.  

From eq. (\ref{eqcont3}) or (\ref{eqfinale}) one can see that to extract the susceptibility of the CNT from the spectra, one first needs to determine the depolarization parameters $f$, $f_y$ and $\delta_0$. One way to access them independently from the CNT spectra is to measure the extinction ratio for each wavelength away from the nanotube (fig \ref{figure5} a). To that purpose, we recorded the variation of the intensity $I_0$ as a function of the angle $\delta$ with the spectrometer. Assuming that $(\delta-\delta_0)$, $f$ and $f_{y}$ are all much smaller than 1, $I_0$ (given by the denominator of eq (\ref{eqfinale})) can be written as :

\begin{equation}
I_0(\omega) = |E_r(\omega)|^2 \left( (\delta-\delta_0(\omega))^2 + f_y(\omega)^2 + f(\omega)^2 \right)
\end{equation}

\noindent where all the parameters are assumed to be dependent of the frequency $\omega$. The intensity for each wavelength is fitted separately (fig 5 inset) to extract three parameters: $f_t^2 = f_{y}^2 + f^2$ from the extinction ratio (figure \ref{figure5}b), $\delta_0$ from the angular deviation (figure \ref{figure5}c) and $|E_r|^2$ which corresponds to the spectrum with polarizers in parallel configuration. These measurements shows the excellent extinction ratio of our system in spectroscopy mode which is in the range of 0.2-0.8$.10^{-4}$. One can note that both coefficients $f_t^2$ and $\delta_0$ display periodic oscillations that inform about the depolarization of the light in the setup: the periodic oscillations are still observed for measurements away from the tube thus showing that the oscillations are not related to the CNT physics but to the setup characteristics, i.e depolarization features. On the one hand, the depolarization parameters $f_{y}$ and $f$ cannot be extracted independently using this characterization so that their relative influence cannot be quantitatively assessed. 
On the other hand, the periodic oscillations of $\delta_0$ with the wavelength evidence coherent depolarization by the optics since $\delta_0$ directly derives from the coherent depolarization (eq. (12) and (13)). Theoretically, such periodic oscillations of $\delta_0$ and $f$ with $\lambda$ can be accounted for by considering that the optics act as a Michelson device with a coherent light depolarization (see SI part 3).

\begin{figure*}
\begin{center}
\includegraphics[scale=0.8]{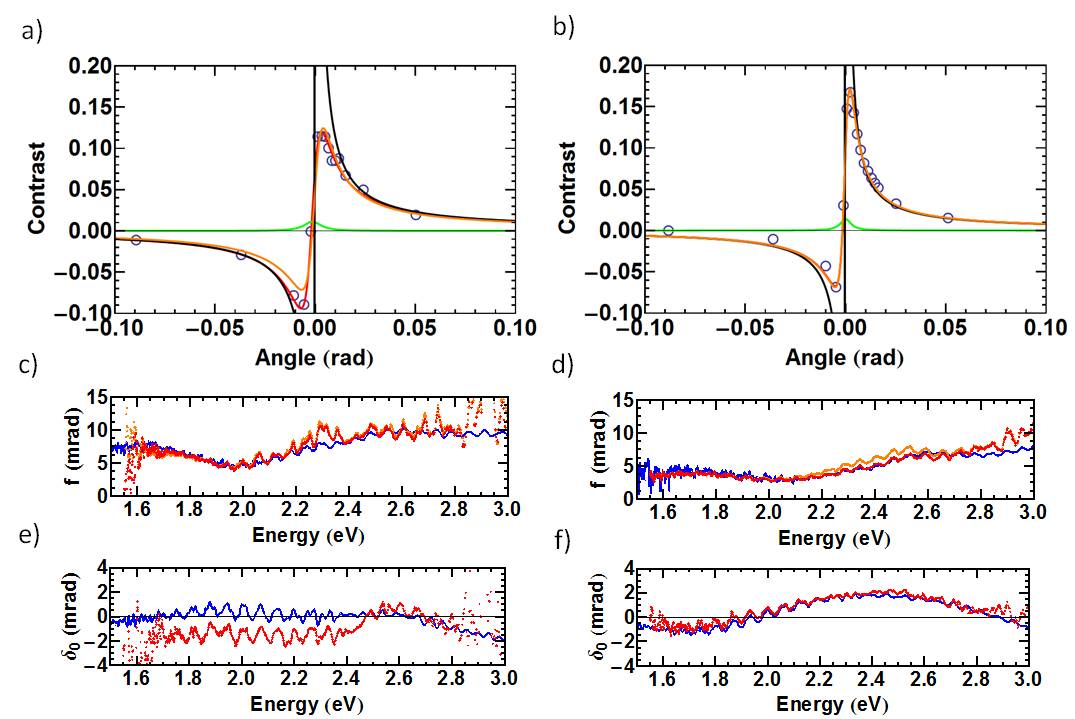}
\end{center}
\caption{a) (resp. b) Contrast evolution of the tubes in case A (resp. B) for one energy corresponding to an absorption peak was selected. Black line : fit from the ideal optics model. Orange line : fit from the non-coherent depolarization model. Red line : fit from the complete model. Green line : scattering term extracted from the complete model. c-d) depolarization parameter $f_t$ extracted from the fits (orange and red: non-coherent and complete model resp.) compared to the the data extracted beside the tube (blue). e-f) $\delta_0$ from the complete model (red) compared to the values measured beside the tube (blue)} 
\label{figure6}
\end{figure*}

\begin{figure*}[ht!]
\begin{center}
\includegraphics[scale=0.8]{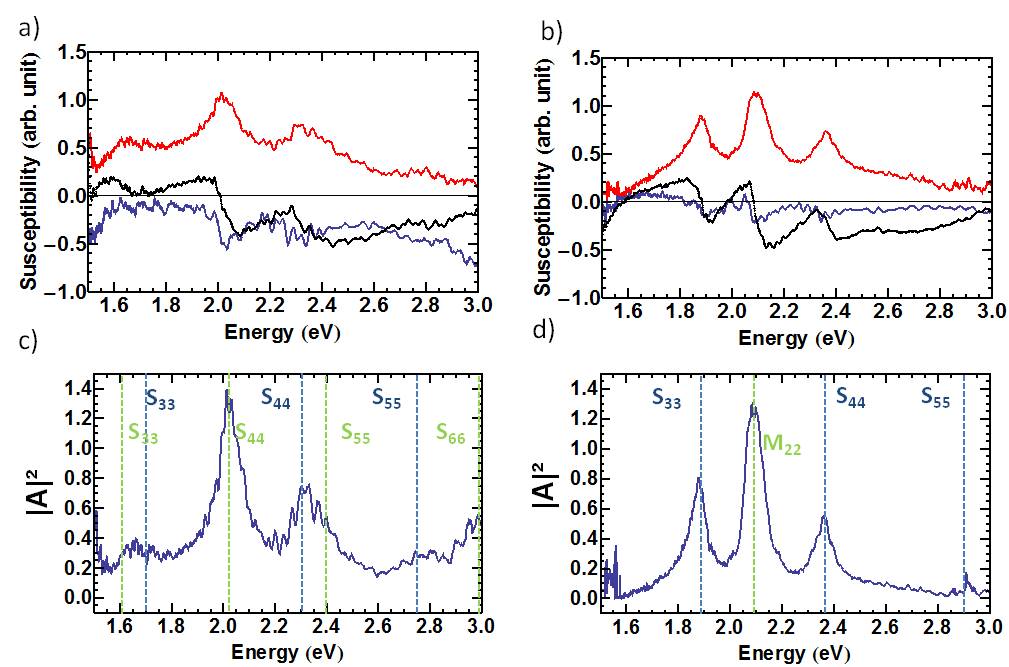}
\end{center}
\caption{a) (resp.b) )Susceptibility extracted for case A (resp. B) :$Re(A)$ (blue line) and $Im(A)$ (red line) extracted from the complete model. The black line represent $Re(A)$ obtained from Kramers-Kronig transformation. c) (resp.d) ) $|A|^2$ extracted from the complete model for case A (resp. case B). The indexation of the resonances is described in the text.} 
\label{figure7}
\end{figure*}

We can now apply the proposed model to fit the experimental nanotube spectra. CNT contrast curves have been measured as a function of the angle $\delta$ following the same protocol as in the previous part. Figures \ref{figure6}a and b show the evolution of this contrast for the tubes in cases A and B (open circles) and for a wavelength corresponding to a maximum of the contrast (i.e an optical transition). The data are fitted using equation \ref{eqfinale} in three cases: ideal optics  ($f=\delta_0=f_y=0$, black line), non coherent depolarization ($f=\delta_0=0$, orange line), complete model with $f$, $\delta_0$ and $f_y$ different from 0. The green line displays the scattering term, extracted from the complete model which is shown to point out the domain of relevance of this term: if one works with angles below $10^{-2}$ rad, this term is not negligible and must be considered in the model. This threshold value for the scattering term only depends on the depolarization characteristics of the setup. If the depolarization by the setup is large, one can neglect the scattering term. One can note that having a non negligible scattering term (as in figures \ref{figure6}a and b) is an evidence of a good polarization conservation by the optical setup.
As expected, the model with ideal optics diverges from the experimental data for small angles $\delta$ which confirms the need for a more realistic model which includes depolarization. Fitting with the non-coherent depolarization model is done by letting all parameters free except the oxide thickness which is known (in this case 100 nm). For the complete model with both types of depolarization, the large number of parameters usually precluded the convergence of the fit with all parameters free. So for illustration purposes, we assumed that the variation of the coherent depolarization parameter $f$, effectively a wave plate effect, was much faster than the non-coherent one $f_y$, which does not possess any phase component, and fixed the latter to a constant value, here $10^{-3}$. This value cannot physically be fixed higher than the minimum value of $f_t$ measured in figure \ref{figure5}, but can be anywhere below it. As a matter of fact, we observed that comparable results are obtained whatever the value chosen for $f_{y}$.\\
As shown in figures \ref{figure6}a and b, the non-coherent model can satisfactorily fit the experimental data in some cases but the complete model usually provides the best fit and is required to satisfactorily fit the data in all experimental cases. Figures \ref{figure6}c and d (resp. e and f) display the coefficient $f_t$ (resp. $\delta_0$) extracted from the fit and plotted as a function of the wavelength for two experimental cases. One can observe that both models provide values of $f_t$ reproducing very well the oscillations which were observed beside the tube (see figure \ref{figure5}). Therefore, this criterion alone cannot allow one to distinguish between the two models. In contrast, the complete model provides $\delta_0$ values which accurately reproduce the variations independently measured beside the tube (see Figures \ref{figure6}e,f) while this parameter is intrinsically  constant with the non-coherent model.\\

To further conclude on the relevance of both models, we plotted the two last parameters $Im(A)$ and $Re(A)$, i.e the real and imaginary parts of the susceptibility using the non coherent model and the complete model. Figures \ref{figure7}a and b shows the results for the complete model as it gave the best agreement. For excitonic optical transitions, it is expected that $Im(A)$ is composed of lorentzian peaks, corresponding to absorption transitions. In the common interpretation that the contrast spectrum is an absorption spectrum \cite{Liu2013}, one would expect that $Im(A)$ would look similar to the black curve on figure \ref{figure2}a. On that criterion, the complete model appears to be much stronger to extract $Im(A)$ than the non coherent model, which proved to be the case for every tube studied. It is also possible to compare $Re(A)$ extracted from the experimental data with the one obtained from Kramers-Kronig transformation (eq \ref{KK}). One can observe a qualitative agreement of the resonances observed in both cases (see black and blue curves in Figures \ref{figure7}a,b), thus providing an additional support for the complete model. The remaining differences probably originate from the current limitations of the model.  First, the fitting procedure could be optimized so that the $f_y$ parameter is let free. Second, as can be seen in figures \ref{figure7}a and b, $Im(A)$ and $Re(A)$ still present some oscillations despite the application of the model with both depolarization types. To better address these oscillations, the model may be complexified by integrating a depolarization Michelson effect although adding more parameters may lead to convergence issues.\\
Extracting both $Im(A)$ and $Re(A)$ also enables one to reconstruct $|A|^2$ which is proportional to $|\chi|^2$ (Figures \ref{figure7}c and d), that is the Rayleigh spectrum of the nanotube. One can see that the spectrum of both figures presents the most intense peaks already visible in the raw spectra of figures \ref{figure2}a and have a flat background, i.e without the influence of the substrate. In particular, the substrate used (100 nm SiO$_2$/Si) induces a minimum of contrast at 400 nm and new features that were hidden on the spectra can now be observed.  The signal-to-noise ratio is also strongly improved since the spectrum combines the data at all tilt angles.\\

This refined optical spectrum can be used to assist and consolidate the structural assignment of carbon nanotubes by Raman spectroscopy as detailed in section 4 of the SI. For instance, after treatment, the optical spectrum in case A (fig. 7c) now displays weaker resonances beside the two main resonances at around 2.0 and 2.3 eV which were already visible. Based on this information, Raman spectroscopy was directly performed at a laser energy (2.33 eV) close to the strong resonance at 2.3 eV. This yielded a semiconducting-type G-band profile and a RBM at 128 cm$^{-1}$ corresponding to a diameter of 1.9-2.0 nm. Using a Kataura plot adapted for nanotubes on SiO$_2$ yields four possible chiralities, all being SC tubes of type 2 \cite{Zhang2015,Petit2012}. Beside a S$_{44}$ resonance at ~2.3 eV, such tubes display resonances at around 1.7 eV (S$_{33}$) and 2.75 eV (S$_{55}$): resonance features are actually observed at these energies thus consolidating the interpretation. It follows that a second nanotube must be considered to account for the resonance at 2.0 eV. Raman spectroscopy was performed at a close excitation energy (1.96 eV) which yielded a semiconducting-type G-band profile with a small G- shift (about 8 cm$^{-1}$ below G+) but no RBM: these observations agree with a type-2 SC tube with diameter of 2.2-2.3 nm (i.e. with RBM at the limit of our experimental range). Beside a S$_{44}$ transition at ~2.0 eV, such tubes have resonances at ~1.6 eV (S$_{33}$), ~2.4 eV (S$_{55}$) and ~3 eV (S$_{66}$). Again, resonance features at these energies are observed after the treatment, thus confirming and consolidating the interpretation of the Raman data.\\
In case B, the refined optical spectrum displays three intense resonances at 1.88 eV, 2.09 eV and 2.36 eV whose profile is much more defined after treatment, together with a smaller peak at  around 2.9 eV. The peaks at 1.88 eV and 2.36 eV display the asymmetric shape characteristic of Rayleigh scattering peaks, thus supporting contribution from a single resonance. The peak at 2.09 eV is more symmetric, which points toward a splitting or more than one contribution. This information was used to directly perform Raman spectroscopy at 2.33, that is close to 2.36 eV. Two low-frequency peaks at 154 cm$^{-1}$ and 118 cm$^{-1}$ were observed, together with a G band made of the strongly downshifted contributions of a semiconducting tube and strongly upshifted contributions of a metallic tube. All these features suggest a DWCNT made of a metallic outer tube and a semiconducting inner tube. The resonances at 1.88 eV, 2.09 eV and 2.36 eV can therefore be assigned to, respectively, the S$_{33}$ of the inner tube, the M$_{22}^-$/M$_{22}^+$ of the outer tube, and the S$_{44}$ of the inner tube. As detailed in SI, this information combined with a model of intertube coupling can be used to interpret the low-frequency modes as coupled breathing modes and refine the number of possible chiralities\cite{Popov2004,Popov20042,Tran2017,Levshov2011}: this yields different DWCNTs but all with an inner SC tube of type 2 with diameter of ~1.7-1.9 nm and an outer M tube with diameter of 2.4-2.55 nm. For the outer M tube, the small splitting between M$_{22}^-$ and M$_{22}^+$ (< 50 meV) points toward a near-armchair M tube (e.g. (20,17)). Such tubes are expected to display an M$_{33}$ resonance at around 2.9 eV. For the inner SC tube, an S$_{55}$ transition would also be expected in the range of 2.7-3 eV (depending on the energy shift induced by the intertube coupling \cite{Liu2014} and the substrate). The weak transition ~2.9 eV may therefore corresponds to these two transitions. Note that the fact that the resonance peaks are much less intense at the extremes of our energy range originates from the effect of the 100 nm SiO$_2$ layer which amplifies the signal in the middle of the range but reduces it at its borders.

\section{Conclusion}

We observed that the polarized optical spectra of individual carbon nanotubes are experimentally more complex than previously reported. They notably display different background shapes depending on the substrate and periodic fluctuations when working at very high extinction ratio. To account for these observations, we developed a model including both coherent and non-coherent depolarization by the optics and the anti-reflection effect of the substrate. Importantly, we showed that the optical response of the substrate cannot be simply removed from the experimental spectra due to its coupling with the complex nanotube susceptibility and the coherent depolarization by the optics. We developed an experimental protocol to measure the depolarization parameters and showed that coherent depolarization is needed to correctly fit all experimental spectra: it allows to reproduce the wavelength-dependence of both the extinction ratio and the angle of maximum extinction $\delta_0$ which would not be possible considering only non-coherent depolarization. Knowing the depolarization factors, our model allows to separate the intrinsic nanotube features from the contribution of the substrate and of the imperfect optics in the experimental spectra. Even though experimental details have to be improved, the proposed model theoretically allows to extract both the real and imaginary parts of the nanotube susceptibility even on substrates with an antireflection layer such as standard SiO$_2$/Si. The method may notably be improved by reducing the coherent depolarization by the optics, improving the fitting algorithm and/or explicitly considering a depolarizing Michelson effect in the model.

\section{Acknowledgments}

Acknowlegments. We thank Xiaoping Hong, Christophe Raynaud and Christophe Voisin for fruitful discussions. This work has been carried out thanks to the support of the LabEx NUMEV project (ANR-10-LABX-20) funded by the «Investissements d’Avenir» French Government program, managed by the French National Research Agency (ANR), the ANR grant Nanophoresis (ANR-13-JS10-0004) and the Federal Grant FA9550-17-1-0027 of the EOARD of the US AFOSR. V. J. acknowledges the support of the Institut Universitaire de France and of the Région Languedoc-Roussillon. Raman spectroscopy was performed at platform SIMS of the University of Montpellier

\section{Supporting Information}

Details on the models mathematical development and on the Raman measurements.

%%%%%%%%%%%%%%%%%%%%%%%%%%%%%%%%%%%%%%%%%%%%%%

%bibliography

%%%%%%%%%%%%%%%%%%%%%%%%%%%%%%%%%%%%%%%%%%%%%%

\end{document}

\bibliographystyle{plain}

\author{L\'{e}onard Monniello}

\author{Huy-Nam Tran}

\author{Rémy Vialla}

\author{Guillaume Prévot}

\author{Said Tahir}

\author{Thierry Michel}

\author{Vincent Jourdain}
\email{vincent.jourdain@umontpellier.fr}

\affiliation{Laboratoire Charles Coulomb (L2C), Université de Montpellier, CNRS, Montpellier, France}

\title{Supplementary information : A comprehensive model of the optical spectra of carbon nanotubes on substrate by polarized microscopy}

\maketitle

The purpose of this document is to complete the article with details of the calculation and complementary experimental results. In the first part we go through the depolarization model in detail. In the second part we detail the optical response from the substrate. In the third part, we address the questions of the periodic oscillations by introducing a depolarizing Michelson interferometer. Finally, we show complementary experimental results.

\section{Depolarization models}

\begin{figure}[h!]
\begin{center}
\includegraphics[scale=0.3]{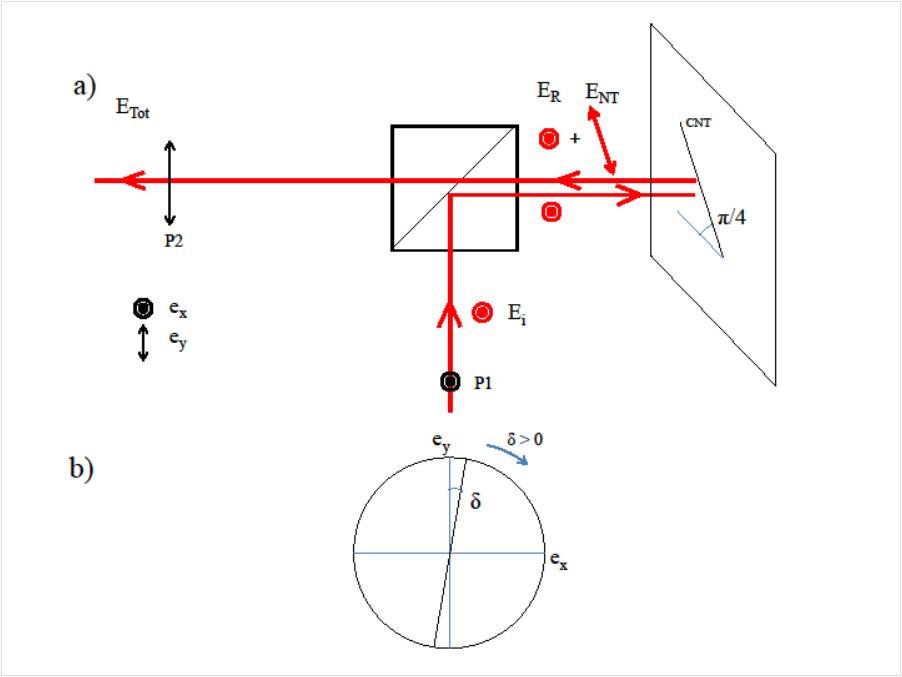}
\caption{\label{schema} Scheme of the polarization-resolved experiment. a) Optical path. The first polarizer (P1) defines the excitation polarization of the field \textit{$E_i$} as being along \textit{$e_x$}. The field is scattered by the substrate without change of the polarization (\textit{$E_R$}) and by the CNT oriented at $\pi /4$ (\textit{$E_{NT}$}). b) The outgoing field is then analyzed by the second polarizer (P2) set at a small angle $\delta$ with the \textit{$e_y$} axis. The arrow shows the convention chosen here for $\delta$>0.}
\end{center}
\end{figure}

In this section, we will show in detail the demonstrations used to obtained the equations for the depolarization models presented in the article following the different steps of the main article. All notations used are similar to the ones of the article. The readers are referred to the main text for the definitions which are not repeated in the SI file. As a reminder, figure \ref{schema} shows a scheme of the experiment and defines the orientation of the axes.

	\subsection{Ideal optical components - no depolarization}

First, let us consider that the optical components are all ideal, and that the polarization is preserved when the light goes through the objective and beam splitter used for the experiment. Figure \ref{schema} shows a schema of the experiment and the definition of the axes:

\begin{itemize}
\item The first polarizer (P1) is set along the $\overrightarrow{e_x}$ axis, thus defining the orientation of the incident field $\overrightarrow{E_i}$

\begin{equation}
\overrightarrow{E_i}=|E_0|e^{i\phi}\overrightarrow{e_x}
\end{equation}

\item The incident light is reflected by the substrate ($\overrightarrow{E_R}$) which does not change the axis of polarization and by the CNT ($\overrightarrow{E_{NT}}$). The field is then defined as $\overrightarrow{E_R}+\overrightarrow{E_{NT}}$ with 

\begin{equation}
\overrightarrow{E_R}=r\overrightarrow{E_i}e^{i\pi}=r|E_i|e^{i\phi}\overrightarrow{e_x}
\end{equation}
\begin{equation}
\overrightarrow{E_{NT}} = A(1+r)^2\frac{|E_i|e^{i\phi}}{2}\left(\overrightarrow{e_x}+\overrightarrow{e_y}\right)
\end{equation}

\item After the second polarizer (P2) considered as ideal, the polarization of the light is set to a small angle ($\delta<<1$) with $\overrightarrow{e_y}$. The amplitude of the field then reads:

\begin{equation}
E_{Tot} = \left(\overrightarrow{E_R}+\overrightarrow{E_{NT}}\right).\left(\sin(\delta)\overrightarrow{e_x}+\cos(\delta)\overrightarrow{e_y}\right)
\end{equation}

\end{itemize}

The small angle approximation leads to $\sin(\delta)\approx\delta$ and $\cos(\delta)\approx1$. We also consider the initial amplitude $E_i$ as equal to 1.

This leads to the following expression of the amplitude of the field as a function of $\delta$:

\begin{equation}
\begin{aligned}
E_{Tot} &= r\delta e^{i\pi} + \frac{Ae^{i\phi}(1+r)^2(1+\delta)}{2}\\
		&\approx r\delta e^{i\pi} + \frac{Ae^{i\phi}(1+r)^2}{2}
\end{aligned}
\end{equation}

\noindent The intensity that is transmitted through $P2$ then reads:

\begin{equation}
\begin{aligned}
I&=|E_{Tot}|^2 = |r\delta e^{i\pi} + \frac{Ae^{i\phi}(1+r)^2}{2}|^2 \\
& \approx r^2\delta^2 + \frac{|A|^2(1+r)^4}{4}+ r(1+r)^2\delta Re\left(e^{i(\pi-\phi)}A^*\right)
\end{aligned}
\end{equation}

The contrast is determined by substracting and normalizing by a reference intensity from the substrate alone $I_0=r^2\delta^2$ and we find:

\begin{equation}
\begin{aligned}
C &=\frac{I-I_0}{I_0}\\
&\approx \frac{|A|^2(1+r)^4}{4r^2\delta^2}+\frac{(1+r)^2Re(e^{i(\pi-\phi)}A^*)}{r\delta}
\end{aligned}
\end{equation}

If $\phi=\pi/2$, then the contrast reads:

\begin{equation}
C \approx \frac{|A|^2(1+r)^4}{4r^2\delta^2}+\frac{(1+r)^2 Im( A)}{r\delta}
\end{equation}

\noindent and we find the expression of equation (6) of the main article.

	\subsection{Depolarization induced by the optical elements}

In this section, we will detail the demonstration of the two models that include depolarization by the optics. In the first part, we consider that the induced depolarization is non-coherent, i.e no phase relation exist in the unpolarized field. In the second part we consider that the optical element preserve the coherence by changing the linear polarization into an elliptic polarization.

	\subsubsection{Non-coherent depolarization}

Unpolarized light can be seen as the sum of two orthogonal linearly (or circularly) polarized fields with random phases:

\begin{equation}
\begin{aligned}
\overrightarrow{E}(t) & = E_x (t) \overrightarrow{e_x} + E_y (t) \overrightarrow{e_y}\\
& = E_x e^{i\psi_x(t)}\overrightarrow{e_x} + E_y e^{i\psi_y(t)} \overrightarrow{e_y}
\end{aligned}
\end{equation}

\noindent where $\psi_x(t)$ and $\psi_y(t)$ are the phases of the $x$ and $y$ components of the field and $t$ is time.
To model this, let us consider that these phases behave like a white noise which can be described by a Langevin Force (see appendix \ref{Annexe A} for more details). Note that with this definition of the unpolarized field, one can check that the intensity of two uncoherently polarized fields will add and not interfere, as expected. \\
The incident field on the sample now reads:

\begin{equation}
\overrightarrow{E_i} = |E_i| \left(a\overrightarrow{e_x}+f_xe^{i\psi_x(t)}\overrightarrow{e_x}+f_ye^{i\psi_y(t)}\overrightarrow{e_y}\right)
\end{equation}

\noindent where $f_x$ (resp. $f_y$) is the proportion of the field with random phase along the $x$ axis (resp. $y$ axis) and $a$ is the proportion of the field that remains polarized. Then for each $t$, $a$ is defined such as 
\begin{equation}
|a+f_xe^{i\psi_x(t)}|^2+|f_ye^{i\psi_y(t)}|^2=1
\end{equation}

With this definition of the random phases, the time average of this equation reads:
\begin{equation}
\langle|a+f_xe^{i\psi_x(t)}|^2+|f_ye^{i\psi_y(t)}|^2\rangle_t=a^2+f_x^2+f_y^2=1
\end{equation}

Then, as previously, the collected field is written $\overrightarrow{E_R}+\overrightarrow{E_{NT}}$ with:

\begin{equation}
\begin{aligned}
\overrightarrow{E_R} & = r|E_i|e^{i\pi}\left(a\overrightarrow{e_x}+f_xe^{i\psi_x(t)}\overrightarrow{e_x}+f_ye^{i\psi_y(t)}\overrightarrow{e_y}\right)\\
\overrightarrow{E_{NT}} &= \frac{A(1+r)^2e^{i\phi}}{2}\left(a + f_xe^{i\psi_x(t)}+f_ye^{i\psi_y(t)}\right)\left(\overrightarrow{e_x}+\overrightarrow{e_y}\right)
\end{aligned}
\end{equation}

The collected intensity and the reference intensity are then equal to:

\begin{equation}
\begin{aligned}
I &=\langle|\left(\overrightarrow{E_R}+\overrightarrow{E_{NT}}\right).\left(\delta \overrightarrow{e_x} + \overrightarrow{e_y}\right)|^2\rangle_t\\
I_0 &= \langle|\overrightarrow{E_R}.\left(\delta \overrightarrow{e_x} + \overrightarrow{e_y}\right)|^2\rangle_t
\end{aligned}
\end{equation}

\noindent Separating the incoherent parts of the expression of $I$ and neglecting $\delta$ over 1, we find:

\begin{equation}
\begin{aligned}
I &= \langle|a\left(ire^{i\pi}\delta+\frac{(1+r)^2Aa}{2}e^{i\phi})\right)\\
&+f_xe^{i\psi_x(t)}\left(ire^{i\pi}\delta+\frac{A(1+r)^2e^{i\phi}}{2}\right)\\
&+f_ye^{i\psi_y(t)}\left(ire^{i\pi}+\frac{A(1+r)^2}{2}e^{i\phi}\right)|^2\rangle_t\\
&= \langle|E_1 + E_2e^{i\psi_x(t)} + E_3e^{i\psi_y(t)}|^2\rangle_t\\
&=|E_1|^2 + |E_2|^2+ |E_3|^2
\end{aligned}
\end{equation}

\noindent where $E_1$, $E_2$ and $E_3$ are the separated parts of the field with no phase. 

Let us call $I_{id}$ (resp. $I_{0id}$) the intensity for the ideal optics case (resp. the reference intensity). After a few steps, we find:

\begin{equation}
\begin{aligned}
I &= I_{id}(a^2+f_x^2)+f_y^2(|\frac{A(1+r)^2}{2}+ri|^2)\\
  &= I_{id}(a^2+f_x^2)+f_y^2(I_{id}+r^2)\\
  &= I_{id} + f_y^2r^2
\end{aligned}
\end{equation}
\begin{equation}
\begin{aligned}
I_0 &= r^2 (a^2\delta^2+f_x^2\delta^2+f_y^2)\\
    &= r^2 (\delta^2(a^2+f_x^2+f_y^2)+f_y^2(1-\delta^2))\\
    &= r^2 (\delta^2 + f_y^2(1-\delta^2)\\
    &\approx I_{0id}+f_y^2r^2
\end{aligned}
\end{equation}

The expression of the contrast is now given by:

\begin{equation}
\begin{aligned}
C&=\frac{I-I_0}{I_0} = \frac{I_{id}}{I_{0id}+f_y^2r^2}\\
&= \frac{|A|^2(1+r)^4}{4r^2\left(\delta^2+f_y^2\right)} + \frac{\delta(1+r)^2 Re(e^{i(\pi-\phi)}A^*)}{r\left(\delta^2 + f_y^2\right)}
\end{aligned}
\end{equation}

\noindent which correspond to equation (8) of the main article.
 
	\subsubsection{Coherent depolarization}

In this second part, we will consider that the depolarization induced by the optical element is coherent, meaning that the linear polarization of the input becomes elliptic. To model this, let us consider that the input field can be written as:

\begin{equation}
\overrightarrow{E_i} = |E_i|(\overrightarrow{e_x}+f_{co}e^{i\psi}\overrightarrow{e_y})
\end{equation}

\noindent with the elliptic polarization being written as the sum of two cross polarizations with a relative phase $\psi$. $f_{co}$ is the proportion of the light that is cross-polarized and we consider $f_{co}<<1$. The total field collected by the microscope can be written as $\overrightarrow{E_R}+\overrightarrow{E_{NT}}$ with:

\begin{equation}
\overrightarrow{E_R} = r|E_i|e^{i\pi}\left(\overrightarrow{e_x}+f_{co}e^{i\psi}\overrightarrow{e_y}\right)
\end{equation}
\begin{equation}
\overrightarrow{E_{NT}} = \frac{(1+r)^2Ae^{i\phi}|E_i|}{2}(\overrightarrow{e_x}+\overrightarrow{e_y})
\end{equation}

\noindent For further calculations, let us take $|E_i|=1$. Then we find that:

\begin{equation}
\begin{aligned}
E_{Tot} &= \left(\overrightarrow{E_R}+\overrightarrow{E_{NT}}\right).\left(\sin(\delta)\overrightarrow{e_x}+\cos(\delta)\overrightarrow{e_y}\right)\\
&\approx r e^{i\pi}\left(\delta+f_{co} e^{i\psi}\right)+\frac{(1+r)^2Ae^{i(\pi-\phi)}}{2}(1+\delta)
\end{aligned}
\end{equation}

The reference field is the reflected part from the sample, and its intensity is $I_0=|re^{i(\pi-\phi)}|^2|\delta +f_{co} e^{i\psi}|^2$. Taking $\phi=\pi/2$, we can write $I-I_0$, which reads:

\begin{equation}
\begin{aligned}
I-I_0 &= \frac{|A|^2}{4} + r Re\left[i(\delta + f_{co}e^{i\psi})A^*\right]\\
&=\frac{|A|^2}{4} + r\left[\delta Im(A) + f_{co} Re(ie^{i\psi}A^*)\right]
\end{aligned}
\end{equation}

Then for any dephasing besides $\psi=0 [\pi]$, and contrary to the two previous cases, the second term of this expression has a component proportional to $Re(A)$. Finally, considering $\delta,f_{co}<<1$, the contrast reads:

\begin{equation}
\label{mirror_coh}
\begin{aligned}
C&=\frac{I-I_0}{I_0} \\
&= \frac{|A|^2}{4r^2|\delta +f_{co} e^{i\psi}|^2} \\
&+ \frac{(\delta+f_{co}\cos\psi)Im(A)- f_{co}\sin\psi Re(A)}{r|\delta +f_{co} e^{i\psi}|^2}
\end{aligned}
\end{equation}

\noindent which corresponds to equation (12) of the main article.

\textit{NOTE : If we consider the first expression of the field as normalized, i.e $\overrightarrow{E_i} = |E_i|(a\overrightarrow{e_x}+f_{co}e^{i\Psi}\overrightarrow{e_y})$ with $$a^2+|f_{co}e^{\Psi)}|^2=1,$$ we obtain a similar expression after the approximation $f_{co},\delta<<1$:
$$C=\frac{|A|^2}{4r^2|a\delta +f_{co} e^{i\psi}|^2}+\frac{r Re\left[i\delta A^* + af_{co}e^{i\psi}A^*\right]}{r^2|a\delta +f_{co}e^{i\psi}|^2}$$ for $a\approx1$, we find the same expression than equation (22). This normalization is only useful in the case of a very strong coherent depolarization, where $f_{co}$ cannot be neglected compared to $a$.}

\section{Nanotubes on silicon oxide: antireflection effect}

In this part we look in detail on the effect of the substrate. We no longer consider only mirror type substrate (fused silica for instance) but more complex ones such as SiO$_2$/Si with different oxide thicknesses. By modifying the thickness, one can change the optical response of the substrate.
The first section of this part is dedicated to the theoretical calculation. In the second part, the calculations are confronted to experimental data.

	\subsection{Effective field on the nanotube}
	
		\subsubsection{Anti-reflection layer}

\begin{figure}
\begin{center}
\includegraphics[scale=0.4]{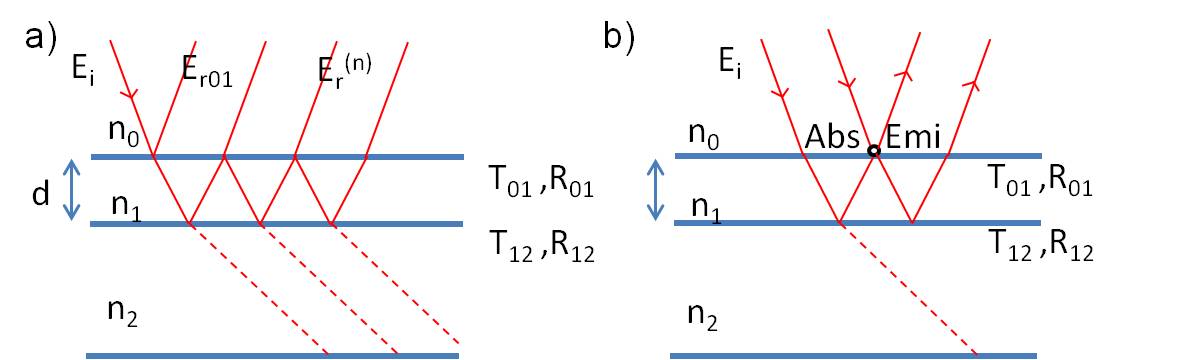}
\caption{\label{FigureS2} a) Scheme of the SiO$_2$/Si substrate and of the transmitted/reflected beams at the two interfaces giving rise to interference. A large angle of incidence has been used for clarity purpose. The first reflected field, on the oxide surface is noted $E_{r01}$. The other reflections $E_r^{(n)}$ come from multiple reflections between the $0-1$ and the $1-2$ interfaces, with $n$ the number of reflections. $R$ and $T$ are, respectively, the reflection and transmission coefficients at each interface. b) Scheme of the local field interacting with the CNT.}
\end{center}
\end{figure}

		The expression of the field reflected by the silicon oxide layer on Si is obtained by the calculation of the Fabry-Perrot interference. As shown in fig. \ref{FigureS2} a), the substrate can be modeled as a layer of index $n_1$ and thickness $d$ on top of a semi-infinite medium of index $n_2$. The reflected field is the sum of two contributions:
	
	\begin{itemize}
		\item $E_{r01}=R_{01}e^{i\pi}E_i$ is the reflected field at the interface between air and the oxide; 
		\item $E_r^{(n)}=T_{01}^2R_{12}^nR_{01}^{n-1}e^{i(n\Phi + n\pi)}E_i$ is the reflected field after $n$ reflections inside the oxide layer;
\end{itemize}		
		
\noindent where $R_{ij}$ and $T_{ij}$ are, respectively, the reflection and transmission coefficients of the interface between medium $i$ and $j$, $\Phi=2\pi (2d)/\lambda$ is the phase acquired through the oxide and the $\pi$ dephasing comes from the reflection at the $0-1$ and $1-2$ interfaces as $n_2 > n_1>n_0$. The values of $R_{ij}=|(n_i-n_j)/(n_i+n_j)|$ are here taken positive as the $\pi$ dephasing is taken into account independently. The expression of the total reflected field is then given by:

\begin{equation}
\begin{aligned}
E_r &= E_{r01} + \sum_{n=1}^{\infty}E_r^{(n)}\\
& = E_i(-R_{01}+\sum_{n=1}^{\infty} T_{01}^2R_{12}^nR_{01}^{n-1}e^{i(n\Phi + n\pi)})\\
&=E_i(-R_{01}+\frac{T_{01}^2}{R_{01}}\sum_{n=1}^{\infty} R_{12}^nR_{01}^{n}e^{i(n\Phi + n\pi)})\\
&=E_i(-R_{01}+\frac{T_{01}^2}{R_{01}}\frac{R_{12}R_{01}e^{i(\Phi + \pi)}}{1-R_{12}R_{01}e^{i(\Phi + \pi)}})\\
&=E_i(-R_{01}-\frac{T_{01}^2}{R_{01}}\frac{R_{12}R_{01}e^{i\Phi}}{1+R_{12}R_{01}e^{i\Phi}})
\end{aligned}
\end{equation}

Knowing that $R^2+T^2=1$ if the medium absorption is negligible, then we obtain the expression:

\begin{equation}
E_r=-E_i \frac{R_{01}+R_{12}e^{i\Phi}}{1+R_{12}R_{01}e^{i\Phi}}
\end{equation}

We see that the oxide layer has an effective reflection coefficient that is complex and depends on its thickness and the wavelength. In a first approximation, one can consider that the indexes of the oxide and silicon are constant over the visible range and $n_1 = 1.47$ and $n_2=4$. We then obtain the reflection coefficient $R_{E}$:

\begin{equation}
R_{E} = -\frac{0.19+0.46e^{4i\pi d/\lambda}}{1+0.09e^{4i\pi d/\lambda}}
\end{equation}

This expression is the one used in the main article to model the effective reflection coefficient.
		
		\subsubsection{Absorption and emission: local field}

\begin{figure}[hb!]
\begin{center}
\includegraphics[scale=0.3]{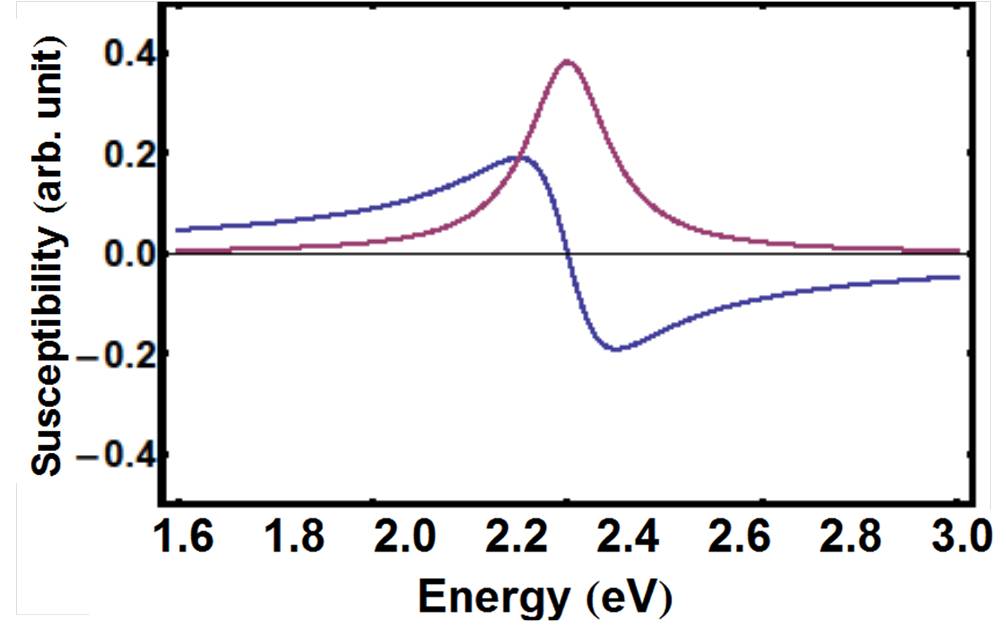}
\caption{\label{FigureS3} Real part and imaginary part of the susceptibility as defined in eq \ref{eq_A}.}
\end{center}
\end{figure}

	In the case of emission and absorption by a nanotube (or any particle) at the interface, the local field seen by the nanotube is composed of both the incident light and the reflected light. In the case of a mirror substrate, this leads to an amplification factor of 1+r as already described in the main text. Figure \ref{figureS2} b) shows a simplified scheme for a layered substrate (for instance SiO$_2$/Si). In such a case, the amplification acts twice: once for the absorption and once for the emission (or scattering). Compared with the effective reflection coefficient, the main difference is that the incident light is not reflected by the substrate meaning that the phase of the field is not changed. The previously destructive interference now becomes constructive. By a similar calculation to previously, we can write the expression of the local field $E_{LO}$ (for Local Oscillator) seen by the tube:

\begin{equation}
\begin{aligned}
\overrightarrow{E_{LO}} &= \overrightarrow{E_i} + \overrightarrow{E_i}\sum_{n=1}^{\infty} T_{01}R_{12}^nR_{01}^{n-1}e^{i(n\phi+n\pi)}\\
&= \overrightarrow{E_i}\left(1-\frac{(1-R_{01}^2)R_{12}e^{i\phi})}{1+R_{01}R_{12}e^{i\phi}}\right)\\
&=\overrightarrow{E_i}R_a
\end{aligned}
\end{equation}

\noindent where $R_a$ is the amplification factor. In the emission (or scattering) case, calculations lead to the same expression. The expression of the scattered field is given by :

\begin{equation}
E_{NT}=E_i(R_a)^2\frac{A}{2}
\end{equation} 

\noindent where the $1/2$ factor comes from the $\pi/4$ orientation of the nanotube with respect to the axes of polarization.
To illustrate the effect on the absorption spectrum in the main text, we consider that the nanotube presents an absorption peak characteristic of excitonic absorption \cite{Berciaud2010} and centered at the energy $\omega_0$. The linear optical susceptibility can be calculated using density matrix formalism and reads \cite{Sebaeian}:

\begin{equation}
\label{eq_A}
A(\lambda) = \frac{C}{(\omega_0-\omega)-i\frac{\Gamma}{2}}
\end{equation}

\noindent where the absorption profile , i.e the imaginary part, is a Lorentzian, $\Gamma$ is the width of the Lorentzian profile and $C$ is a constant that depends on the optical properties of the absorbing element such as electric dipole  moment or carrier density. Figure \ref{FigureS3} shows the real part and the imaginary part of the susceptibility as a function of the wavelength $\lambda$ for $C= 3$ and $\Gamma=25$ nm.

	\subsection{Experimental validation on different substrates.}

\begin{figure}
\begin{center}
\includegraphics[scale=0.6]{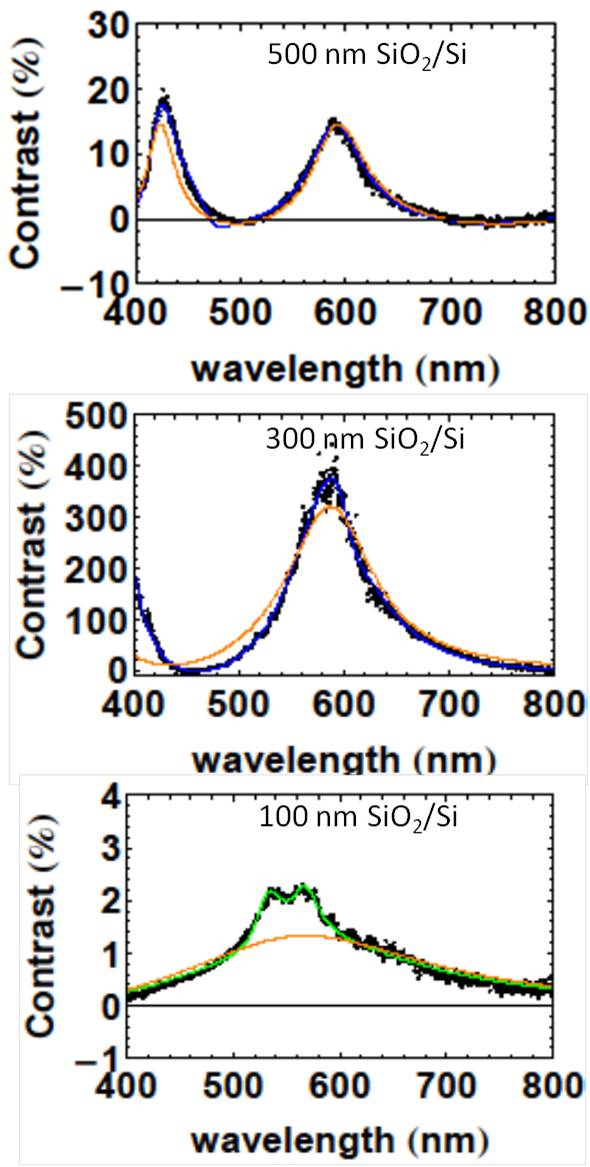}
\caption{\label{substrats} Measured contrast for substrate with different oxide thickness: 100 nm, 300 nm and 500 nm. Measures were performed on bundles or coils showing an high contrast. red curves are fits using only the substrate reflectivity with a non resonant absorption. Blue curves and green curve are fits using two lorentzian profiles in addition of the non resonant absorption.}
\end{center}
\end{figure}

	To observe the influence of the substrate, we measured the contrast spectra of nanotubes for three different thickness of $SiO_2$: 100 nm, 300 nm and 500 nm.
Figure \ref{substrats} shows the contrast measured on those substrate. For 300 nm and 500 nm thicknesses, the substrate effect is expected to show rather sharp peaks so, in order to enhance the broad signal, we realised those spectra on bundles of nanotubes displaying an intense signal. The idea was to add all the contributions of the non resonant absorption while having a lower contribution from the lorentzian resonant profiles. Data (black) have first been fitted using only the non resonant contribution (red curve). While the agreement between the fit and the data is already good, we also performed the fit adding two lorentzian profiles. This would be supported by the assumption that the tubes within the bundle are all within the same range of diameters and thus display peaks that are close to each other. We observe a perfect match for all the fits (blue and green curve), thus validating the model of the substrate reflectivity used in the paper.

\section{Depolarizing Michelson interferometer}

	In this part, we propose to improve the model of depolarization by taking into account internal reflections inside the cube beam-splitter. This aims to explain the periodic oscillations that are observed both on the spectra for small angle $\delta$ and in the extracted value $f$ and $\delta_0$.
	
	In this model we consider two paths for the light beam. The first one has already been described in the first part of the this Supplementary information and in the main article: the light is reflected by the cube into the microscope objective, interacts with the sample and is detected by the spectrometer. The detected intensity $I_0$, as already described in eq. (20) of the main article, writes as:
	
\begin{equation}
I_0=|E_i|^2r^2\left((\delta-\delta_0)^2+f^2\right)
\end{equation} 

Let us consider now that the light beam is also reflected inside the cube beam-splitter. The cube acts as a Michelson interferometer and the detected intensity $I_m$ without polarization consideration would be given by:

\begin{equation}
I_m = |E_i|^2 r_c^2 (1 + A \cos \frac{B}{\lambda})
\end{equation}

\noindent where $r_c$ is the reflection coefficient of the cube interface, $A$ is the overlap between the two reflected beams inside the cube and $B=2\pi n \Delta x$ is the optical length difference between the two beams. We now consider that a part of the light is also coherently depolarized by the beam-splitter and the intensity can thus be written as:

\begin{equation}
I_m=|E_i|^2 r_c^2 (1 + A cos \frac{B}{\lambda} )((\delta-\delta_1)^2 + f_2^2)
\end{equation}

\noindent where $\delta_1$ and $f_2$ are the depolarization factors related to the beam inside the cube and are supposed different from $\delta_0$ and $f$ respectively.
These two intensities are detected by the same spectrometer but do not interfere as the white source does not have a long coherence time. 
The total detected intensity is then given by:

\begin{equation}
\begin{aligned}
I_{tot}&=I_0+I_m\\
&=|E_i|^2\left(r^2((\delta-\delta_0)^2+f^2)+r_c^2((1+A cos \frac{B}{\lambda})+f_2^2)\right)
\end{aligned}
\end{equation}
	
to simplify this expression, we define two terms:

\begin{equation}
\alpha=r_c^2(1+A \cos \frac{B}{\lambda})
\end{equation}	
\noindent the intensity fluctuation induced by the michelson, and

\begin{equation}
f_3^2=f^2+\alpha f_2^2
\end{equation}

\begin{figure}
\begin{center}
\includegraphics[scale=0.6]{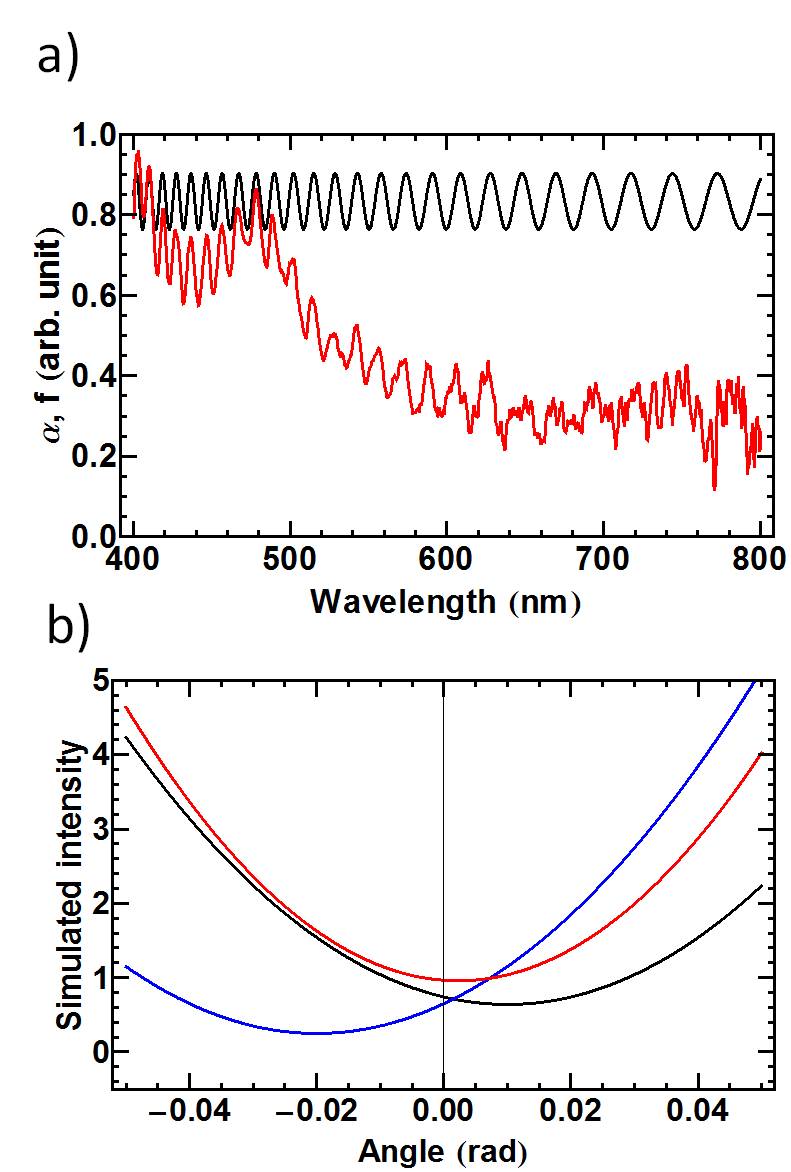}
\caption{\label{mich} a) black: $\alpha$ coefficient plotted with respect to the wavelength using parameters A=1/10 and B=125 $\mu$m. Red curve is $f$ from main text shown for comparison purpose. b) blue (resp. black) Simulated intensity in the model without Michelson effect with parameters $\delta_0$=0.1 (resp. -0.2) and $f$=0.5 (resp. 0.8). Red: simulated intensity with the Michelson effect showing the combined effect of both functions.}
\end{center}
\end{figure}

$\alpha$ is plotted in Figure \ref{mich} a) (black curve) as a function of the wavelength. The parameters A and B used to define the function have been chosen manually to match the relative amplitude and the frequency of the oscillation of the variations of $f$ as has been measured in the main article (red curve). Here, the overlap coefficient is A=1/10 and the optical length difference is B=125 $\mu$m. The total intensity can be written again as:

\begin{equation}
I_{tot}=|E_i|^2\left((\delta-\delta_0)^2 + \alpha(\delta-\delta_1)^2 +f_3^2\right)
\end{equation}

\noindent where we can see that $f_3$ acts as an effective depolarization factor that is modulated by $\alpha$. The depolarization angle is now different from $\delta_0$ and $\delta_1$ and is also modulated by $\alpha$. Figure \ref{mich} b) shows the simulated intensity with respect to the angle of $I_{black}=(\delta-0.1)^2+0.5^2$, $I_{blue}=(\delta+0.2)^2-0.8^2$, i.e intensity evolutions without a Michelson effect and $I_{red}=(I_{black} + \alpha*I_{blue}$ calculated at the wavelength $\lambda$=500 nm and with the same parameters A and B as previously. Because of the nature of the dependence on $\alpha$, the red curve is highly dependent on the wavelength.
We show here that the sum of the functions gives a second order polynom with different values of $\delta_0$ and $f$. This model therefore allows to qualitatively and qualitatively reproduce the experimental periodic fluctuations. However, adding it to the model would leave too many free parameters for the the fitting algorithm to converge and would therefore make the model unpractical to use.

\section{Combined Raman and optical spectroscopy studies}
	
This section is dedicated to show the results of the Raman measurements. Figures S6 to S9 shows the study on the tubes of case A and figures S10 to S13 shows the study on the tubes of case B.

\null\newpage
\makeatletter\onecolumngrid@push\makeatother

\begin{figure*}[h!]
\includegraphics[scale=1]{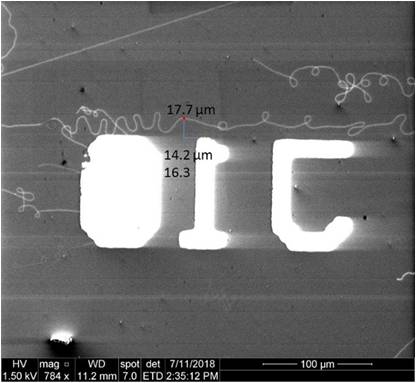}
\caption{SEM image of case A. The red dot indicates the position where the optical and Raman spectra were measured.}

\includegraphics[scale=1]{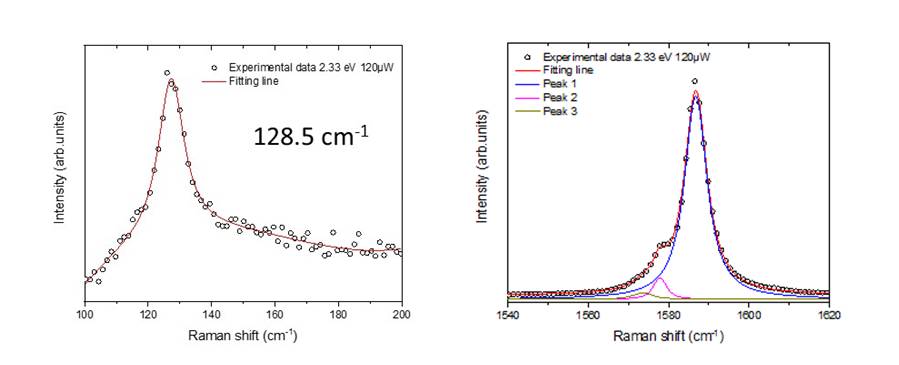}
\caption{RBM and G band of case A recorded at an excitation energy of 2.33 eV. }

\includegraphics[scale=1.3]{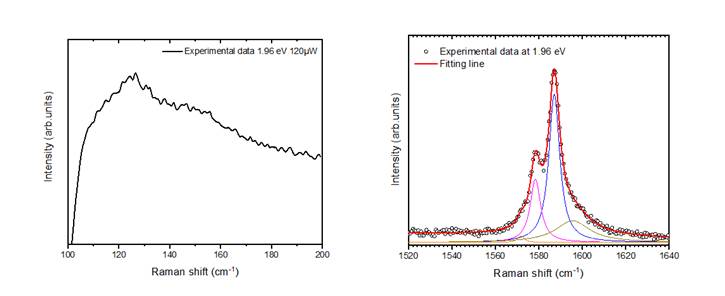}
\caption{RBM and G band of case A recorded at an excitation energy of 1.96 eV.}
\end{figure*}

\clearpage

\begin{figure*}[h!]
\begin{center}
\includegraphics[scale=1.3]{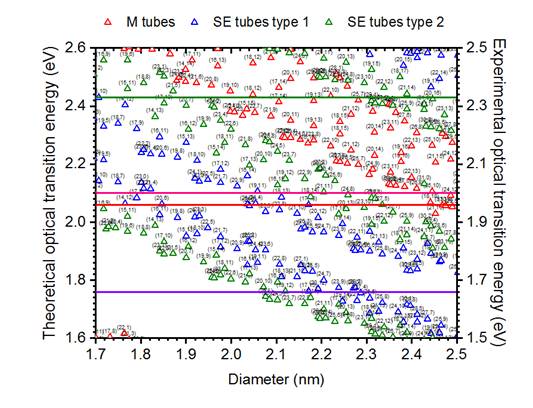}
\caption{Kataura plot indicating the resonance energies in the studied range of diameter and energy. On the right side axis, the experimental transition energies were downshifted by 100 meV to account for the effect of the SiO$_2$ substrate.}
\end{center}
\end{figure*}

\clearpage

\begin{figure*}[h!]
\begin{center}
\includegraphics[scale=1]{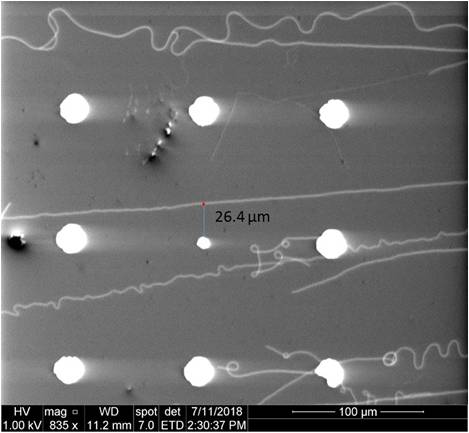}
\caption{ SEM image of case B. The red dot indicates the position where the optical and Raman spectra were measured.}

\includegraphics[scale=0.5]{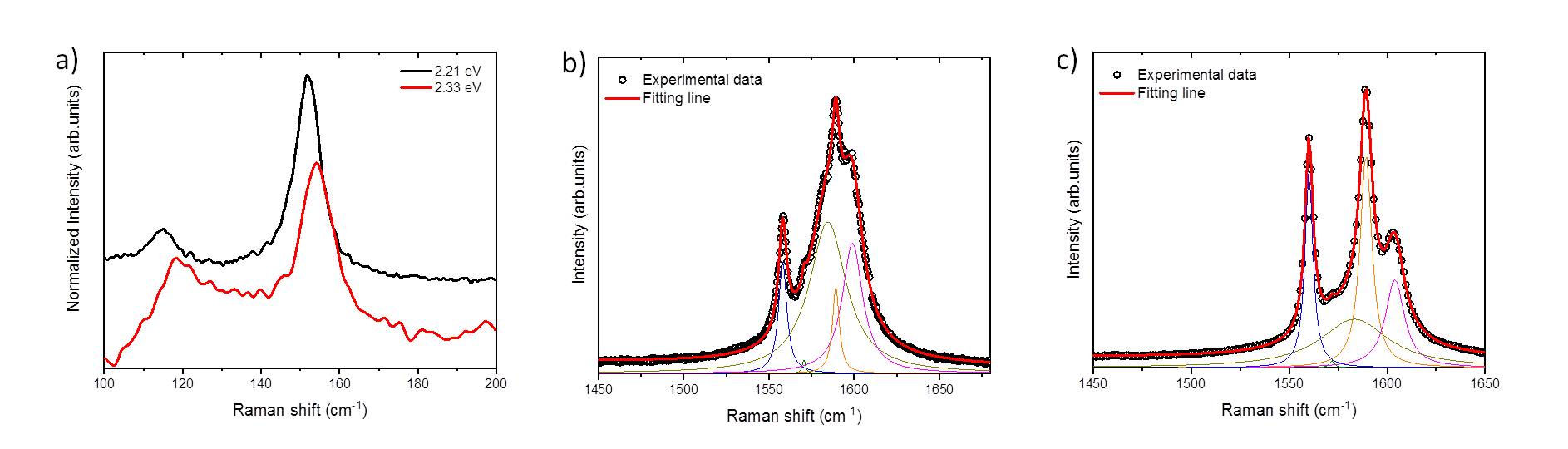}
\caption{RBM and G band of case B. a) RBM recorded at the excitation energies 2.21 and 2.33 eV., b) and c) G band recorded at an excitation energy of 2.21 eV and 2.33 eV respectively.}

\includegraphics[scale=1.2]{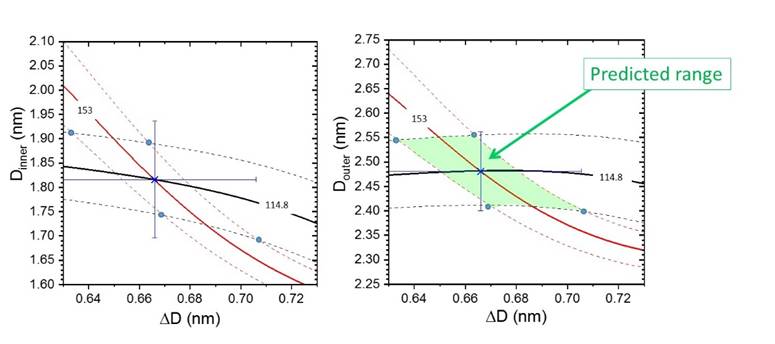}
\caption{DWCNT modeling with RBLM\cite{Tran2017} of inner tube at 153 cm-1 and RBLM of outer tube at 114.8 cm$^{-1}$}
\end{center}
\end{figure*}

\begin{figure*}[h!]
\begin{center}
\includegraphics[scale=1.3]{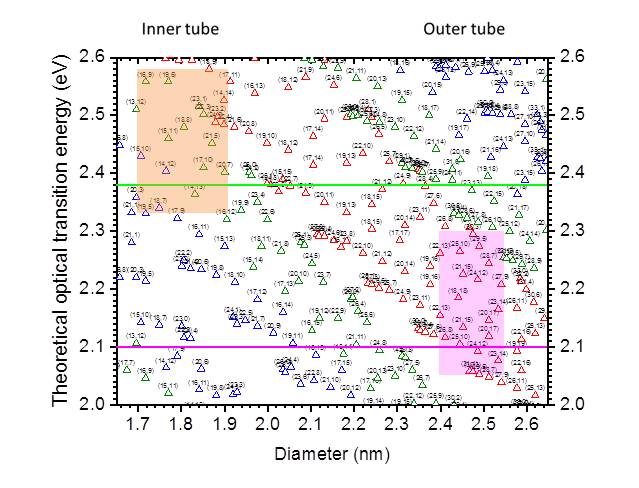}
\caption{Kataura plot indicating the resonance energies in the studied range of diameter and energy.\cite{Popov2004, Popov20042}.}
\end{center}
\end{figure*}

\clearpage
\makeatletter\onecolumngrid@pop\makeatother

\appendix
 \renewcommand{\theequation}{A\arabic{equation}}
\section{Langevin force and non coherent polarization}
\label{Annexe A}

In the first part of this document, we are interested in modeling the depolarization induced by the microscope objective. In the case of non-coherent depolarization, i.e a loss of polarization, the light can be described as composed of the initial polarized light and a non polarized light. In this appendix, we will focus on modeling this non polarized light. 

As already mentioned, non polarized light can be seen as two orthogonal linearly (or circularly) polarized lights with random phases between the two directions of polarization:

\begin{equation}
\begin{aligned}
\overrightarrow{E}(t) & = E_x (t) \overrightarrow{e_x} + E_y (t) \overrightarrow{e_y}\\
& = E_x e^{i\psi_x(t)}\overrightarrow{e_x} + E_y e^{i\psi_y(t)} \overrightarrow{e_y}
\end{aligned}
\end{equation}

\noindent where $\psi_x(t)$ and $\psi_y(t)$ are the phase of the $x$ and $y$ components of the field. If the field is non polarized, then those phases are random and no phase relation can be written.

To model this, let us consider that those phases behave like a white noise and use the Langevin force $F_i$ to describe their fluctuations. This force is commonly used to model the Brownian motion of particles, or any other stationary stochastic process \cite{Bylander} with delta-function correlation,

\begin{equation}
\left\langle F_i(t)F_j(t')\right\rangle = \Gamma\delta_{i,j}\delta(t-t')
\end{equation}
	
\noindent where the angular brackets indicate statistical averaging, $\lbrace i,j\rbrace = \lbrace x,y \rbrace$ and $\delta$ is the Kronecker delta symbol showing that the evolution of the two phases is completely uncorrelated. $\Gamma$ is the amplitude of the correlation function. The phase $\psi_i$ is then given by:

\begin{equation}
	\psi_i(t) = \int_0^t F_i(u)du
\end{equation}

\noindent The correlation on the phase gives the relations:

\begin{equation}
\langle\psi_i(t)\psi_j(t')\rangle = \Gamma (t+t'-|t-t'|)\delta_{i,j}\\
\end{equation}

\noindent And, in the stationary limit:

\begin{equation}
\begin{aligned}
\langle e^{i\psi(t)} \rangle & = e^{-\Gamma t} \rightarrow 0\\
\langle e^{i\psi_i(t) - i\psi_j(t')} \rangle &= e^{-\Gamma |t-t'|}\delta_{i,j}
\end{aligned}
\end{equation}

Let us consider three fields $E_1$, $E_2$ and $E_3$ defined by:
\begin{equation}
\begin{aligned}
E_1(t) &= Ae^{i\psi_1(t)}\\
E_2(t) &= Be^{i\psi_2(t)}\\
E_3(t) &= C \\
\end{aligned}
\end{equation}

Here, the intensity that can be detected is the average over time of the instantaneous intensity:

\begin{equation}
\begin{aligned}
I_1 &= \langle |E_1(t)|^2\rangle_t = A^2\\
I_2 &= \langle |E_2(t)|^2 \rangle_t= B^2\\
I_3 &= \langle |E_3(t)|^2\rangle_t = C^2 \\
\end{aligned}
\end{equation}

Now we can calculate the intensity resulting from the interference between these fields:

\begin{equation}
\begin{aligned}
I_{12} &= \langle |E_1(t)+E_2(t')|^2\rangle_t + \langle |Ae^{i\psi_1(t)}+Be^{i\psi_2(t')}|^2 \rangle_t\\
       &= \langle A^2+ B^2 + (ABe^{i\psi_1(t)}e^{-i\psi_2(t')}+ cc )\rangle_t\\
       &= A^2 + B^2 + \langle (ABe^{i\psi_1(t)}e^{-i\psi_2(t')}+ cc )\rangle_t \\
       &= A^2 + B^2 + 2e^{-\Gamma |t-t'|}\delta_{i,j} = A^2 + B^2 + 0\\
I_{13} &= \langle |E_1(t)+E_3(t)|^2\rangle_t \\
       &= A^2 + C^2 + ACe^{-\Gamma t} \rightarrow A^2 + C^2
\end{aligned}
\end{equation}

where $cc$ stands for conjugate complex. We see here that this definition correspond to the intuitive definition of non coherent lights as two non coherent light will not interfere.

%%%%%%%%%%%%%%%%%%%%%%%%%%%%%%%%%%%%%%%%%%%%%%

%bibliography

%%%%%%%%%%%%%%%%%%%%%%%%%%%%%%%%%%%%%%%%%%%%%%

%%%%%%%%%%%%%%%%%%%%%%%%%%%%%%%%%%%%%%%%%%%%%%

%%%%%%%%%%%%%%%%%%%%%%%%%%%%%%%%%%%%%%%%%%%%%%

%%%%%%%%%%%%%%%%%%%%%%%%%%%%%%%%%%%%%%%%%%%%%%